\documentclass[twocolumn,dvipsnames,prx,nofootinbib,floatfix,superscriptaddress]{revtex4-2}

\usepackage{xcolor}
\usepackage{amsmath,amssymb}
\usepackage{appendix}
\usepackage{makecell}
\usepackage{relsize}
\usepackage{graphicx}
\usepackage{stmaryrd}
\usepackage{bm}
\usepackage{lipsum}
\usepackage{multirow}

\usepackage[frozencache,cachedir=./]{minted}
\setcellgapes{4pt}
\usepackage[caption=false, position=top]{subfig}

\usepackage[colorlinks,plainpages]{hyperref}
\hypersetup{urlcolor=blue,citecolor=blue, linkcolor=blue,hypertexnames=false}

\usepackage{xspace}
\usepackage{setspace}
\usepackage{bm}

\newcommand{\ppop}{\ensuremath{p(\theta|\Lambda)}}
\newcommand{\ppopi}{\ensuremath{p(\theta_i|\Lambda)}}

\newcommand{\pinj}{\ensuremath{p(\theta|\Lambda_\mathrm{inj})}}
\newcommand{\pinji}{\ensuremath{p(\theta_i|\Lambda_\mathrm{inj})}}

\newcommand{\ppe}{\ensuremath{p(\theta|\Lambda_\mathrm{pe})}}

\newcommand{\pdettheta}{\ensuremath{P(\mathrm{det}|\theta)}}
\newcommand{\hatpdettheta}{\ensuremath{\hat{P}(\mathrm{det}|\theta)}}
\newcommand{\hatpdetthetai}{\ensuremath{\hat{P}(\mathrm{det}|\theta_i)}}
\newcommand{\hatpdetthetaj}{\ensuremath{\hat{P}(\mathrm{det}|\theta_j)}}

\newcommand{\pdetlambda}{\ensuremath{P(\mathrm{det}|\Lambda)}}

\newcommand{\KICP}{\affiliation{Kavli Institute for Cosmological Physics, The University of Chicago, 5640 S. Ellis Ave., Chicago, IL 60615, USA}}

\begin{document}


\title{A neural network emulator of the Advanced LIGO and Advanced Virgo selection function}

\author{Thomas A. Callister}
\KICP{}

\author{Reed Essick}
\affiliation{Canadian Institute for Theoretical Astrophysics, 60 St. George St, Toronto, Ontario M5S 3H8, Canada}
\affiliation{Department of Physics, University of Toronto, 60 St. George Street, Toronto, ON M5S 1A7, Canada}
\affiliation{David A. Dunlap Department of Astronomy, University of Toronto, 50 St. George Street, Toronto, ON M5S 3H4, Canada}

\author{Daniel E. Holz}
\KICP{}
\affiliation{Department of Physics, The University of Chicago, Chicago, IL 60637, USA}
\affiliation{Department of Astronomy \& Astrophysics, The University of Chicago, Chicago, IL 60637, USA}
\affiliation{Enrico Fermi Institute, The University of Chicago, Chicago, IL 60637, USA}

\begin{abstract}
    Characterization of search selection effects comprises a core element of gravitational-wave data analysis.
    Knowledge of selection effects is needed to predict observational prospects for future surveys and is essential in the statistical inference of astrophysical source populations from observed catalogs of compact binary mergers.
    Although gravitational-wave selection functions can be directly measured via injection campaigns---the insertion and attempted recovery of simulated signals added to real instrumental data---such efforts are computationally expensive.
    Moreover, the inability to interpolate \textit{between} discrete injections limits the ability to which we can study narrow or discontinuous features in the astrophysical distribution of compact binary properties.
    For this reason, there is a growing need for alternative representations of gravitational-wave selection functions that are computationally cheap to evaluate and can be computed across a continuous range of compact binary parameters.
    In this paper, we describe one such representation.
    Using pipeline injections performed during Advanced LIGO \& Advanced Virgo's third observing run (O3), we train a neural network emulator for $\pdettheta{}$, the probability that a given compact binary with parameters is successfully detected, averaged over the course of O3.
    The emulator captures the dependence of $\pdettheta{}$ on binary masses, spins, distance, sky position, and orbital orientation, and it is valid for compact binaries with component masses between $1$--$100\,M_\odot$.
    We test the emulator's ability to produce accurate distributions of detectable events, and demonstrate its use in hierarchical inference of the binary black hole population.
\end{abstract}

\maketitle

\section{Background}
\label{sec:intro}

Like most astronomical experiments, gravitational-wave astronomy suffers from selection biases: the Advanced LIGO~\cite{Aasi2015} and Advanced Virgo~\cite{Acernese2015} instruments most readily detect compact binary mergers that are massive, nearby, and situated in preferred sky positions and orientations~\cite{Schutz1987,Finn1993,Cutler1994,Flanagan1998}.
Unlike many other experiments, however, these selection biases are almost exactly quantifiable.
The gravitational-wave signatures of merging compact binaries are, in most cases, described by vacuum General Relativity alone and can therefore be calculated from first principles to a high degree of accuracy.
These calculated waveforms can then be injected, either via hardware or software, into real instrumental data and analyzed with search pipelines~\cite{DalCanton2014,Adams2016,Klimenko2016,Usman2016,Messick2017,Sachdev2019,Aubin2021,Drago2021}, allowing us to accurately replicate the survey and directly determine how often signals are missed or found~\cite{abbott_rate_2016,Biwer2017,O2-pop,O3a-pop,O3b-pop}.

Such knowledge of the gravitational-wave selection function is essential to any physical or astrophysical interpretation of gravitational-wave data.
The selection function is needed to forward model and predict future observations, given models for the compact binary population and the physics governing it~\cite[e.g.,][]{Dominik2015,Rodriguez2019,Belczynski2020,Bavera2022,Broekgaarden2022,Fragione2023}.
In the reverse direction, the selection function is a critical ingredient in inference: the reverse-engineering of intrinsic source populations from incomplete and noisy catalogs of detected gravitational-wave signals~\cite{Farr2015,Mandel2019,Vitale2020,O3a-pop,O3b-pop}.

In principle, injection campaigns allow the selection function to be calculated to high precision (though still subject to systematic uncertainties like imperfect detector calibration or imperfect gravitational waveform models). 
In practice, however, we rapidly run into problems of dimensionality and scale.
Because compact binary mergers are described by at least 15 parameters (the components' masses and spins, binary position and orientation, etc.), it is impractical to directly compute detection probabilities for every possible combination of binary properties.
Additionally, processing mock signals injected into data is computationally expensive and labor intensive, requiring at least as much time and person-power as the actual searches for real gravitational waves.
The field therefore relies on several widespread shortcuts to estimate selection functions and detection probabilities.
Inference of the compact binary population, for example, usually relies on a large suite of reference injections, whose properties are randomly drawn from an astrophysically plausible distribution of compact binary parameters~\cite{Tiwari_2018,Farr2019,Essick2021,Essick2022,injections}.
This fixed injection suite can be importance (re)sampled to target other distributions that are sufficiently ``close'' to the chosen reference distribution, and various integrals over the space of detectable binary parameters can be approximated as Monte Carlo averages over this discrete set of injections~\cite{Essick2022,Essick2023}.

There are difficulties with this paradigm, however.
In the forward direction, a fixed set of reference injections cannot be immediately used to quantify the detectability of some new gravitational-wave source not included among the original injections.
In the reverse direction, the use of reference injections in population inference may become computationally intractable as gravitational-wave catalogs continue to grow in size.
In order for systematic uncertainties in search selection functions to remain subdominant, it has been estimated that the number of reference injections must scale at least linearly with the number of detected gravitational-wave events, although sublinear scaling may be achievable with the use of low-discrepancy sequences~\cite{Essick2022,Talbot2023}.

Due to the computational requirements of pipeline injections, a source's anticipated signal-to-noise ratio (SNR) is a common semianalytic proxy for detectability.
In this case, one calculates the optimal SNR $\rho_{\rm opt}$ of a given source, or, preferably, an ``observed'' SNR $\hat \rho$ that includes the effects of random noise fluctuations, and demands that detections exceed some detection threshold $\rho_{\rm thresh}$~\cite{Fishbach2018,Fishbach-most-massive,Gerosa2020,gwmockcat,Essick2023}.
However, although large signal-to-noise ratios are a necessary condition for detection, they are not \textit{sufficient}; gravitational-wave detection additionally relies on auxiliary signal consistency checks not captured by simple 
SNR estimates~\cite[e.g.,][]{Allen2005}, and instrumental noise more readily mimics some types of signals than others.
It is likely that these higher order effects can be semianalytically mimicked by adopting a \textit{variable} SNR threshold that that depends on the given source parameters~\citep{Essick2023}, but further development is needed.

In this paper we describe an alternative paradigm for evaluating and correcting for selection effects in gravitational-wave astronomy.
Specifically, we construct and demonstrate the use of a neural network that is trained to emulate the Advanced LIGO and Advanced Virgo selection function during the most recent ``O3'' observing run~\cite{abbott_open_2023,GWTC3,injections}.
Although trained on a suite of discrete injections analyzed by compact binary search pipelines, the emulator instead learns the latent, continuous function characterizing detection probabilities as a function of compact binary parameters (see Fig.~\ref{fig:cartoon}).
The result is a means of quickly and inexpensively ``interpolating'' between reference injections.
This allows for simple and precise statements regarding the detectability of new gravitational-wave signals and source populations when forward modeling future observations.
In the reverse direction, the emulator offers a scalable and computationally-feasible route towards population inference with ever-growing gravitational-wave catalogs.

The rest of this paper is organized as follows.
In Sec.~\ref{sec:training} we describe the construction and training of our neural network emulator.
In Sec.~\ref{sec:performance}, we then demonstrate the trained network's ability to accurately predict distributions of detected events and as well as integrated detection efficiencies. 
In Sec.~\ref{sec:hierarchical-inference}, we discuss the emulator's use in hierarchical inference.
We carry out and compare inference of the binary black hole population using (\textit{i}) standard injections and (\textit{ii}) the trained emulator.
We conclude in Sec.~\ref{sec:discussion} with a comparison to existing work and a discussion of future applications.

\section{Training a detection probability emulator}
\label{sec:training}

\begin{figure}
    \centering
    \includegraphics[width=0.5\textwidth]{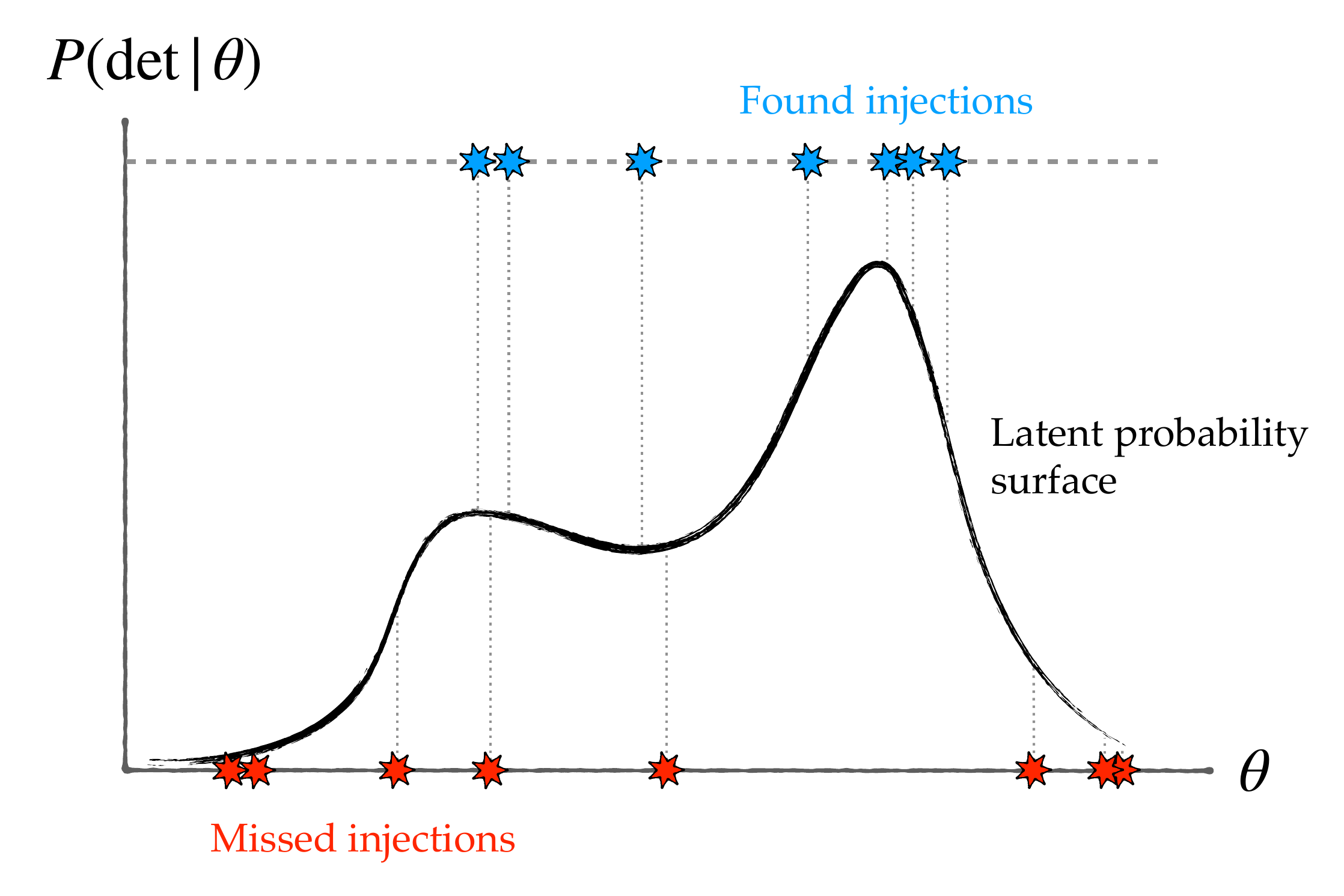}
    \caption{
    A cartoon illustrating the latent detection probability surface we seek to model using machine learning, and the discrete realization of missed and found events that serve as our training data.
    For a given gravitational-wave signal with source parameters $\theta$ and a random time of arrival, there exists some probability \pdettheta{} that it will be successfully detected.
    Knowledge of \pdettheta{} is required both to forward model the detectability of different theorized source populations, and to infer underlying populations from observed catalogs.
    }
    \label{fig:cartoon}
\end{figure}

Our goal is to learn the probability \pdettheta{} that a given gravitational-wave source, with parameters $\theta$, would have been detected if it occurred at a random time during the Advanced LIGO \& Advanced Virgo O3 observing run.
The situation is illustrated in Fig.~\ref{fig:cartoon}.
Abstractly, there exists some surface \pdettheta{} that defines the probability of gravitational-wave detection.
We do not have direct access to this surface, however, but instead only indirect information via the locations of successfully detected (``found'') and undetected (``missed'') mock signals injected into gravitational-wave data with various values of $\theta$~\cite{injections}.
Each such injection formally serves as an individual Bernoulli trial, sampling the underlying \pdettheta{} surface and returning a single success or a single failure.
This process can be repeated at many different times to sample different realizations of instrumental noise and/or periods of terrestrial contamination.

We attempt to learn this latent \pdettheta{} surface using a simple feed-forward artificial neural network.
We adopt a standard multilayer perceptron architecture that takes in a vector of binary parameters $\theta$, parses this input via several densely-connected layers, and concludes with a single output neuron.
The output neuron has sigmoid activation function, yielding an estimated detection probability $\hatpdettheta{} \in \{0,P_\mathrm{max}\}$, where $P_\mathrm{max} = 0.94$ is the approximate fraction of time that one or both of the LIGO-Hanford and LIGO-Livingston detectors were online during O3~\cite{abbott_open_2023}.
This threshold reflects the fact that arbitrarily loud signals would still be missed $6\%$ of the time, arriving when neither detector is active.\footnote{Although this estimate neglects time in which only the Advanced Virgo instrument was online, Advanced Virgo was significantly less sensitive than the Advanced LIGO instruments in O3, and so the error in this approximation is likely small.}
The exact architecture used does not significantly impact the emulator's performance, but we find the best results when adopting four hidden layers with 192 neurons each.

In the following subsections, we describe our training process in more detail, including the training data and loss functions used.
Creating and training even a very simple feed-forward neural network, however, requires a multitude of additional design choices, such as the precise balance of training datasets, activation functions, neuron initialization, and learning rate.
Details like these are described further in Appendix~\ref{app:tuning}.

\subsection{Training Data}

\begin{figure*}
    \centering
    \includegraphics[width=\textwidth]{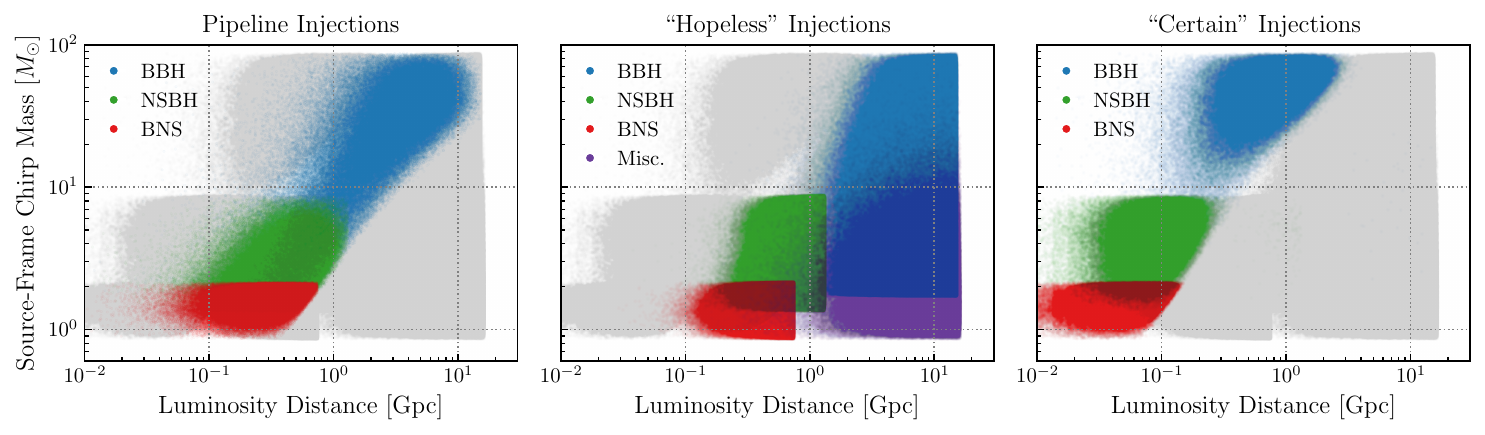}
    \caption{
    Summary of the training data used in constructing our emulator for the Advanced LIGO and Advanced Virgo selection function.
    Shown in the left panel are the source-frame chirp masses and luminosity distances of artificial signals injected into LIGO-Virgo data and processed by compact binary search pipelines to determine which were missed and found.
    Several such populations were produced and analyzed, corresponding to binary black holes (blue), neutron-star black hole binaries (green), and binary neutron stars (red).
    We augment these with additional sets of ``hopeless'' injections, with no chance of successful detection; these are highlighted in the middle panel.
    We additionally train on sets of ``certain'' injections, shown in the right panel, which are nearly certain to be identified provided one or more Advanced LIGO detectors are operational at the time of their arrival.
    Although we display training data in the mass-distance plane here, our emulator is trained on the comprehensive space of binary masses, spins, and extrinsic parameters.
    Note that the ``hopeless,'' ``certain,'' and pipeline injections do not cleanly separate in the chirp mass-luminosity distance plane; this is due to the influence of these other parameters on compact binary detection.
    }
    \label{fig:training}
\end{figure*}

We take as training data the set of publicly-released software injections performed during the Advanced LIGO \& Advanced Virgo O3 observing run~\cite{GWTC3,O3b-pop,injections}.
These comprise simulated gravitational-wave signals added into real O3 data and subsequently passed through several detection pipelines~\cite{DalCanton2014,Adams2016,Klimenko2016,Usman2016,Messick2017,Sachdev2019,Aubin2021,Drago2021}.
Three such injection sets were prepared, representing the populations of intermediate- and stellar-mass binary black holes, neutron star-black hole binaries, and binary neutron stars; the source-frame chirp masses and luminosity distances of these three datasets are highlighted in the left-hand side of Fig.~\ref{fig:training}.
We consider an injection to have been ``detected'' if it was assigned a false-alarm rate (FAR) of less than one per year in at least one search pipeline.
This matches the selection criterion adopted by the LIGO-Virgo-KAGRA Collaboration in its analysis of the compact binary population following O3~\cite{O3a-pop,O3b-pop}.

When the O3 injection sets were constructed, only events with a nontrivial chance of being detected (more precisely, events with expected network signal-to-noise ratios exceeding 6 were retained~\cite{O3a-pop,injections}.
This causes the absence of low-mass injections at large distances in the left-hand side of Fig.~\ref{fig:training}.
Meanwhile, because of the volumetric prior and mass distribution used to generate injections, vanishingly few events are situated at very close distances with large masses.
In order to accurately learn \pdettheta{}, we also need training data representing these very weak and very loud signals.
We therefore augment the O3 injection sets in two ways.
First, we generate several additional injection sets, but this time retain only events that \textit{fail} the network SNR cut (that is, ``hopeless'' events that are certain to be undetectable).
These events, shown in the center column of Fig.~\ref{fig:training}, are all marked as ``missed.''
Second, we generate sets of injections with expected matched-filtering SNRs of at least $20$ in one or both of the Hanford and Livingston detectors.
Shown in the right-hand side of Fig.~\ref{fig:training}, we regard these as ``certain'' detections, effectively guaranteed to be found provided that one or more detectors are operating at the time of their arrival.
We mark $\sim 6\%$ of the certain detections as ``missed.''
As discussed above, this is the percentage of detections that that would arrive when neither neither Hanford nor Livingston were operational~\cite{abbott_open_2023}, and are therefore necessarily unobservable.
The remainder of the certain detections are labeled as ``found.''
See Appendix~\ref{app:training-data} for a further description of these additional injection sets.

\subsection{Defining a loss function}
\label{sec:loss}

We train our \pdettheta{} emulator by maximizing the likelihood of having obtained the recorded outcomes (missed or found) of the injections described above.

Let the set $\{\theta_i\}$ describe the parameters of all injections in our training data, and $\{\lambda_i\}$ be a set of binary flags recording whether each injection was found ($\lambda_i = 1$) or missed ($\lambda_i=0$).
Given a model $\hatpdettheta{}$, the likelihood of having obtained this particular realization of missed and found injections is
    \begin{equation}
    \begin{aligned}
    & p(\{\lambda_i\}|\hatpdettheta{}) \\
        &\qquad= \prod_{{\rm Found}\,i} \hatpdetthetai
            \prod_{{\rm Missed}\,j} \left[1-\hatpdetthetaj\right] \\
        &\qquad= \prod_i \left[\hatpdetthetai\right]^{\lambda_i}
            \left[1-\hatpdetthetai\right]^{1-\lambda_i}.
    \end{aligned}
    \end{equation}
Rather than maximize this likelihood, we find it useful to instead maximize the posterior $p(\hatpdettheta{}|\{\lambda\})$, with non-trivial priors on predicted detection probabilities,
    \begin{equation}
    \begin{aligned}
    &p(\hatpdettheta{}|\{\lambda\}) \\
        &\quad\propto \prod_i e^{-\beta \hatpdetthetai} \left[\hatpdetthetai\right]^{\lambda_i}
            \left[1-\hatpdetthetai\right]^{1-\lambda_i},
    \end{aligned}
    \label{eq:training-posterior-1}
    \end{equation}
for some constant $\beta$.
The parameter $\beta$ functions to disfavor large detection probabilities, particularly in regions of parameter space with sparse training data.
The corresponding loss function is
    \begin{equation}
    \begin{aligned}
    \mathcal{L}(\hatpdettheta{})
        &= -\ln p(\hatpdettheta{}|\{\lambda\}) \\
        &= \sum_i \big[
            \beta \hatpdetthetai
            - \lambda_i \ln \hatpdetthetai \\[-7pt]
            &\hspace{1.5cm} - (1-\lambda_i)\ln(1-\hatpdetthetai)
            \big].
    \end{aligned}
    \label{eq:phat-loss}
    \end{equation}
This is the standard binary cross-entropy loss function commonly used in classification, with the additional $\beta$-dependent penalization.
In the absence of this penalization, our network overpredicts detection probabilities at large distances and low masses where we have few samples; we empirically find that choosing $\beta = 0.35$ alleviates this.

As we will discuss further below, hierarchical inference of the compact binary population does not depend on \pdettheta{} directly, but instead on the \textit{integral} of \pdettheta{} over the full parameter space of compact binaries:
    \begin{equation}
    \xi(\Lambda) \equiv \pdetlambda = \int d\theta \pdettheta{} \ppop.
    \label{eq:xi}
    \end{equation}
Here, $p(\theta|\Lambda)$ is the probability distribution of binary parameters according to some specific model for the compact binary population, denoted $\Lambda$, and $\xi(\Lambda)$ is the fraction of all binaries in this population we expect to successfully detect.
This quantity is also called the \textit{detection efficiency}.
We find that neural networks trained to optimize pointwise detection probabilities, using Eq.~\eqref{eq:phat-loss}, do not necessarily yield good estimates of the integrated detection efficiencies needed for population inference.
We therefore augment the loss function to explicitly penalize poorly recovered detection efficiencies and guide the network towards accurate recovery of $\xi(\Lambda)$.

Prior to training, we randomly draw sets of compact binary parameters, $\{\theta_{I,j}\}$, from several different population models $p(\theta|\Lambda_I)$.
Here, we use $I$ to index population models and $j$ to index the individual draws from a given population.
Within each training step, we compute the detection efficiencies for these reference populations as predicted by the emulator in its current state, each of which can be approximated via the average
    \begin{equation}
    \hat \xi_I \approx \frac{1}{N_I} \sum_j^{N_I} \hat{P}(\mathrm{det}|\theta_{I,j}),
    \label{eq:mc-efficiency}
    \end{equation}
where $N_I$ is the number of draws from population $I$.
For each population, the detection efficiency can also be estimated via reweighting of the pipeline training injections.
If $p(\theta|\Lambda_{\rm inj})$ is the distribution from which these injections were drawn and $N_{\rm total}$ is the total number of injections performed, then
    \begin{equation}
    \xi_I \approx \frac{1}{N_{\rm total}} \sum_{\mathrm{Found}\,i}
        \frac{p(\theta_i|\Lambda_I)}{\pinji},
    \label{eq:xi-reweighting}
    \end{equation}
If we take $\xi_I$ estimated in this manner as our target, then we expect the product $N_I \xi_I$ to be Binomial distributed with mean $N_I \xi_I$ and standard deviation $\sqrt{N_I \xi_I (1 - \xi_I)} \approx \sqrt{N_I \xi_I}$.
In the limit of large $N_I$, the distribution can be well-approximated by a Gaussian, and the likelihood of $\hat \xi_I$ itself is
    \begin{equation}
    p(\hat \xi_I | \xi_I) \propto \exp\Big[ -\frac{\big(\hat\xi_I - \xi_I\big)^2}{2\,\xi_I/N_I}\Big].
    \label{eq:xi-likelihood}
    \end{equation}
Strictly, $\hat \xi_I$ and $\xi_I$ are each noisy estimators of some unknown, underlying detection efficiency, over which we could marginalize.
We find via direct evaluation that this procedure yields only small corrections to Eq.~\eqref{eq:xi-likelihood}.

We use Eq.~\eqref{eq:xi-likelihood} to define an additional loss term
    \begin{equation}
    \begin{aligned}
    \mathcal{L}_\xi
        &= \sum_I -\ln p(\hat \xi_I | \xi_I,N_I) \\
        &= \sum_{I} \frac{\big(\xi_I - \hat \xi_I\big)^2}{2\,\xi_I/N_I}.
    \end{aligned}
    \label{eq:xi-loss}
    \end{equation}
We employ four reference populations in this manner:
the three distributions traced by the binary black hole, neutron star-black hole, and binary neutron star pipeline injections~\cite{injections}, and a fourth population designed to approximate the observed astrophysical population of binary black hole mergers.
Further details are provided in Appendix~\ref{app:reference-populations}.

The total training loss is given by the sum of Eqs.~\eqref{eq:phat-loss} and \eqref{eq:xi-loss}.

\subsection{Parametrization of compact binary mergers}

We take as input a 13-dimensional description of a compact binary: component masses, component spins, distance, sky position, binary inclination, and polarization angle.\footnote{
    We neglect the time and phase of binary coalescence.
    The resulting \pdettheta{} computed by our emulator should therefore be regarded as a time- and phase-averaged detection probability, assuming uniform distributions for each quantity.
    }
Exactly \textit{how} these parameters are presented to the neural network, however, strongly affects network performance.
We find that the best performance is achieved when providing the squared amplitudes
    \begin{equation}
    \begin{aligned}
    \mathcal{A}_+^2 &= \left(\frac{(\mathcal{M}^{\rm det}_c/M_\odot)^{5/6}}{D_L/{\rm Gpc}} \frac{1+\cos^2\iota}{2}\right)^2 \\
    \mathcal{A}_\times^2 &= \left(\frac{(\mathcal{M}^{\rm det}_c/M_\odot)^{5/6}}{D_L/{\rm Gpc}} \cos\iota\right)^2
    \end{aligned}
    \label{eq:amps}
    \end{equation}
as direct inputs to the neural network; these determine, at leading post-Newtonian order, the expected signal-to-noise ratios in ``plus'' and ``cross'' polarizations for inspiral-dominated signals~\cite[e.g.,][]{Finn1993,Buonanno2009}.
Here, $\mathcal{M}_c^{\rm det}$ is the detector-frame binary chirp mass [related to the source-frame chirp mass by $\mathcal{M}^{\rm det}_c = \mathcal{M}_c (1+z)$, where $z$ is the source's redshift], $D_L$ is the luminosity distance, and $\iota$ is the inclination angle between a binary's orbital angular momentum and our line of sight.
We also find it advantageous to ``overspecify'' binary masses, providing the network with both detector-frame total masses and chirp masses, as well as both the standard and symmetric mass ratios (also known as ``data augmentation'' in machine learning settings).
Although only two of these four parameters are needed to fully specify the component masses of a binary, different inspiral stages and waveform effects are more strongly governed by different combinations of these parameters.
We expect that a neural network would therefore need to ``learn'' each of these parameters anyway, a step that is saved if we instead provide them directly.

In contrast, we do not provide the network with all six spin degrees of freedom, but instead compress spin information into several ``effective'' parameters that are expected to capture most inspiral and precessional dynamics.
We include the standard effective inspiral spin~\cite{Racine2008,Santamaria2010,Ajith2011},
    \begin{equation}
    \chi_\mathrm{eff} = \frac{\chi_1 \cos\theta_1 + q \chi_2\cos\theta_2}{1+q},
    \end{equation}
as well as an analogous ``asymmetric'' spin parameter,
    \begin{equation}
    \chi_\mathrm{diff} = \frac{\chi_1 \cos\theta_1 -  \chi_2\cos\theta_2}{2},
    \label{eq:chi-diff}
    \end{equation}
where $\chi_i$ and $\theta_i$ are the dimensionless component spin magnitudes and polar tilt angles.
Both $\chi_\mathrm{eff}$ and $\chi_\mathrm{diff}$ appear in the leading-order spin corrections to the gravitational-wave inspiral phase entering at 1.5PN order~\cite{Kidder1993}.
We also include the generalized precessing spin parameter~\cite{Gerosa2021},
    \begin{equation}
    \begin{aligned}
    \chi^{\rm gen}_{\rm p} = \Big[ 
        &\left(\chi_1 \sin\theta_1\right)^2
        + \left(\tilde \Omega \chi_2\sin\theta_2\right)^2 \\
        &+ 2\tilde \Omega \chi_1\chi_2\sin\theta_1\sin\theta_2\cos\Delta\phi
        \Big]^{1/2},
    \end{aligned}
    \label{eq:chi-p}
    \end{equation}
an extension of the precessing spin parameter first introduced in Ref.~\cite{Schmidt2015}.
Here, $\Delta\phi$ is the angle subtended by the two component spins, once projected onto the plane perpendicular to the orbital angular momentum, and $\tilde \Omega = q (3+4q)/(4+3q)$.

\begin{table}[]
    \footnotesize
    \setlength{\tabcolsep}{4pt}
    \renewcommand{\arraystretch}{1.2}
    \centering
    \caption{
    Compact binary parameters provided to the neural network \pdettheta{} emulator.
    }
    \begin{tabular}{r | l }
    \hline \hline
    Parameter & Definition \\
    \hline
    $\ln\mathcal{A}_+^2$ & Log squared amp. of ``$+$'' polarization; see Eq.~\eqref{eq:amps} \\
    $\ln\mathcal{A}_\times^2$ & Log squared amp. of ``$\times$'' polarization; see Eq.~\eqref{eq:amps} \\
    $\mathcal{M}_c^{\rm det}/M_\odot$ & Detector-frame chirp mass\\
    $M_{\rm tot}^{\rm det}/M_\odot$ & Detector-frame total mass\\
    $D_L/{\rm Gpc}$ & Luminosity distance \\
    $\eta$ & Reduced mass ratio: $m_1 m_2/M_\mathrm{tot}^2$ \\
    $q$ & Mass ratio: $m_2/m_1$ (where $m_2\leq m_1$) \\
    $\alpha$ & Right ascension\\
    $\sin \delta$ & Sine declination\\
    $|\cos\iota|$ & Absolute value of cosine inclination\\
    $\sin\psi$ & Sine of polarization angle\\
    $\cos \psi$ & Cosine of polarization angle\\
    $\chi_\mathrm{eff}$ & Effective inspiral spin\\
    $\chi_\mathrm{diff}$ & ``Antisymmetric'' inspiral spin; see Eq.~\eqref{eq:chi-diff} \\
    $\chi^{\rm gen}_{\rm p}$ & Generalized precessing spin; see Eq.~\eqref{eq:chi-p}\\
    \hline
    \hline
    \end{tabular}
    \label{tab:input-params}
\end{table}

The complete set of parameters passed to the neural network is listed in Table~\ref{tab:input-params}.

\section{Performance of the trained emulator}
\label{sec:performance}

\begin{figure*}
    \centering
    \includegraphics[width=\textwidth]{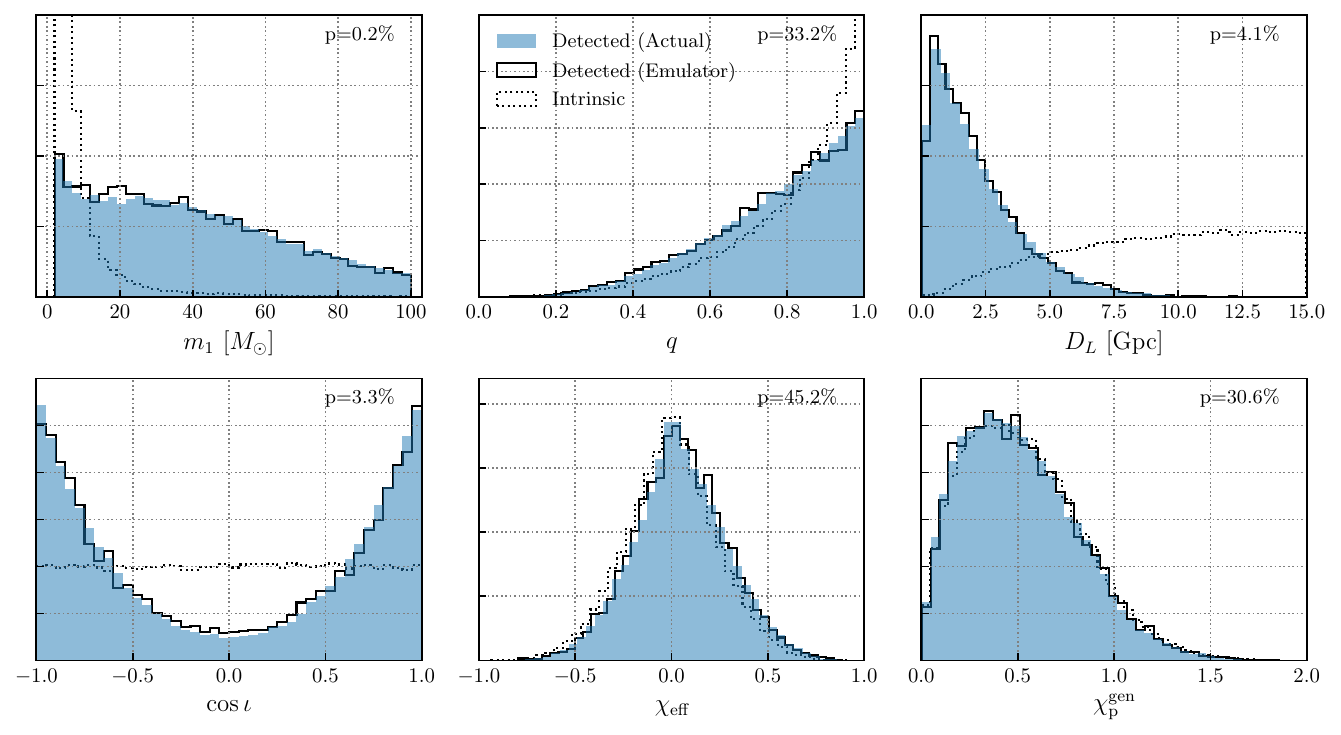}
    \caption{
    Distribution of detected binary black hole mergers as predicted by our trained neural network \pdettheta{} emulator (solid black distributions), compared to the distributions of found pipeline injections (found pipeline injections).
    Both populations are drawn from identical intrinsic distributions (dotted black).
    The trained neural network emulator produces distributions of found binary black holes that are near matches to the actual distribution of found events from compact binary search pipelines.
    The numbers inset in the upper-right corner of each plot show $p$-values of Kolmogorov-Smirnov test statistics between the emulated and actual distributions of found events.
    These $p$-values indicate good statistical agreement between most pairs of distributions, but also that some pairs are not formally indistinguishable.
    The emulated and actual distributions of $m_1$ values, for instance, have a $p=2\times10^{-3}$ probability of being drawn from the same parent distribution.
    }
    \label{fig:bbh}
\end{figure*}

\begin{figure*}
    \centering
    \includegraphics[width=\textwidth]{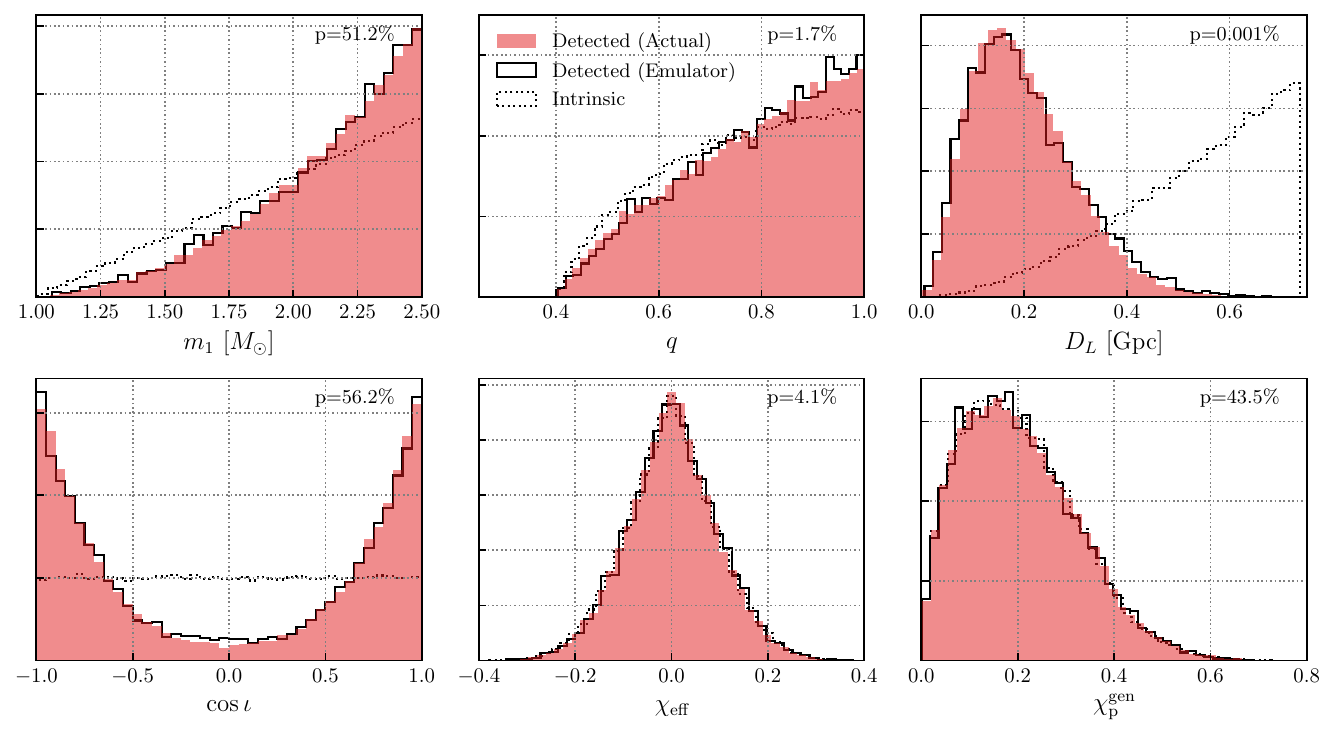}
    \caption{
    As in Fig.~\ref{fig:bbh} but for the population of detectable binary neutron stars.
    }
    \label{fig:bns}
\end{figure*}

\begin{figure*}
    \centering
    \includegraphics[width=\textwidth]{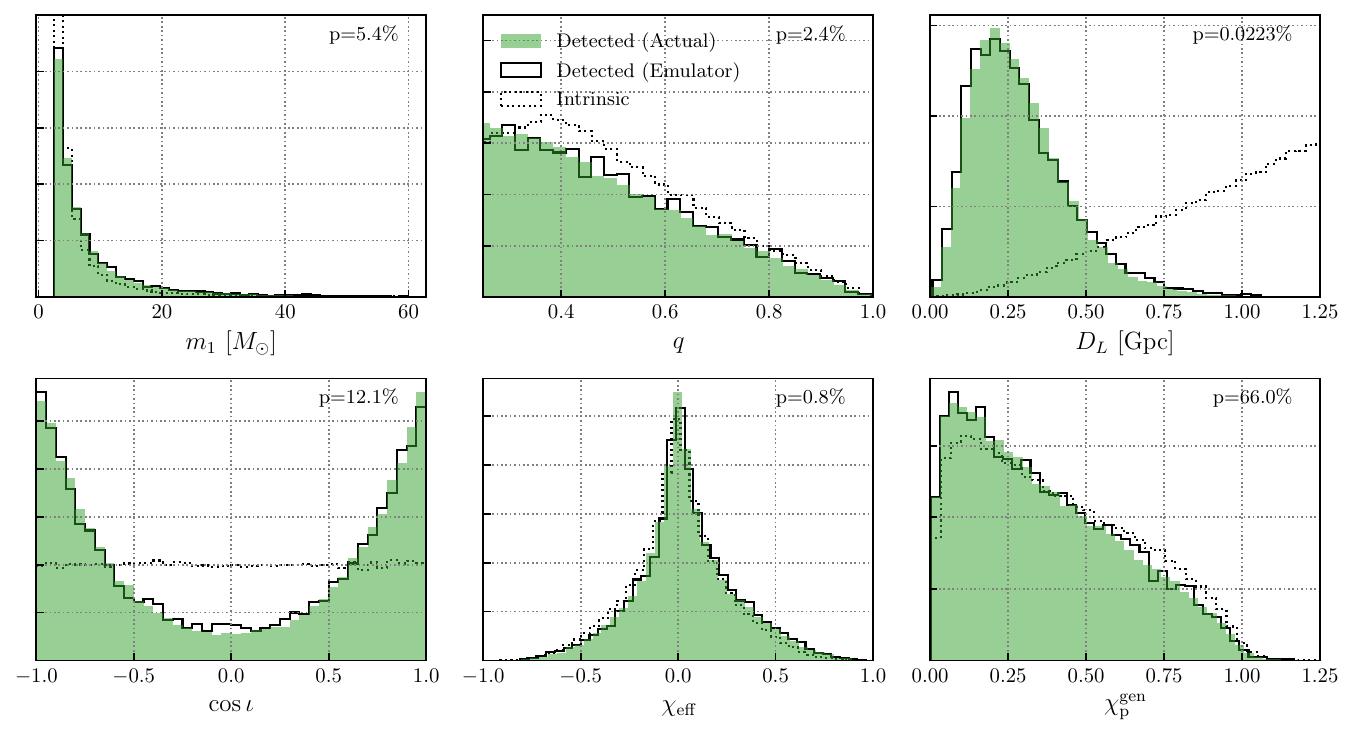}
    \caption{
    As in Fig.~\ref{fig:bbh} but for the population of detectable neutron star-black hole binaries.
    }
    \label{fig:nsbh}
\end{figure*}

\begin{figure}
    \centering
    \includegraphics[width=0.48\textwidth]{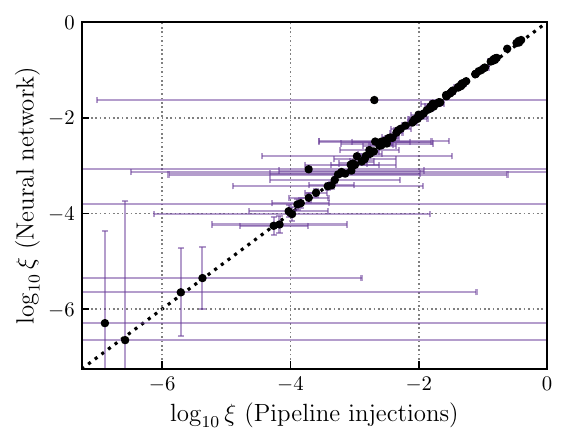}
    \caption{
    Comparison of integrated binary black hole detection efficiencies as computed by traditional reweighting of pipeline injections ($x$-axis) vs. our trained neural network emulator ($y$-axis).
    Each point represents a different possible binary black hole population, with randomly chosen mass, mass ratio, spin, and redshift distributions.
    The error bars show expected uncertainties, given the finite number of pipeline injections/population draws informing each estimate.
    The large majority of proposed populations yield efficiency estimates that are consistent between the two methods.
    There exist a small number of outlier points, for which the methods do not agree; these correspond to populations for which the pipeline injection-based efficiency estimates are highly uncertain, due to poor reweighting efficiencies from the injections' parent distribution.
    }
    \label{fig:efficiency-check}
\end{figure}

Figure~\ref{fig:bbh} demonstrates the performance of the trained network using binary black hole signals.
The blue histograms show the distribution of found events among the binary black hole pipeline injections (see the left-hand panel of Fig.~\ref{fig:training})~\cite{injections}.
The underlying population from which these injections were drawn is indicated via the dotted histograms; this distribution is described further in Appendix~\ref{app:training-data}.
To test the accuracy with which our neural network emulates \pdettheta{}, we draw a new set of proposed binary black holes from this same underlying distribution.
We use our trained network to assign detection probabilities \hatpdettheta{} to each of these systems; these are then rejection sampled in accordance with the predicted probabilities to randomly identify a subset as successfully detected.
This process is repeated until we gather $10^4$ new detections.
The resulting distributions of found events are shown via the solid black histograms.
Figures~\ref{fig:bns} and \ref{fig:nsbh} analogously show the reconstructions of detected binary neutron star and neutron star-black hole binary populations.

The detected distributions of each class of compact binary, as predicted by the neural network emulator, are good visual matches to the distributions of compact binaries actually detected by search pipelines.
As a more quantitative metric, we compute Kolmogorov-Smirnov test statistics between actual (blue) and emulated (solid black) distributions of found injections; the $p$-values of these test statistics are shown in the upper-right corner of each subplot.
Most $p$-values lie above $10^{-2}$, indicating good statistical agreement.
Some $p$-values are lower, though.
The actual and emulated distributions of binary neutron star distances, for example, have only a $10^{-5}$ chance of being drawn from the same parent distribution.

A well-working \pdettheta{} emulator should not only produce the correct distributions of detected compact binary parameters, but must also reproduce the correct \textit{absolute fraction} of events that are successfully detected (the former does not necessarily imply the latter).
The absolute detection efficiency $\xi$, defined above in Eq.~\eqref{eq:xi}, is necessary to successfully predict gravitational-wave detection rates and is a critical ingredient in the statistical inference of astrophysical compact binary populations.
To test the ability of the trained emulator to produce accurate detection efficiencies, we repeatedly and randomly draw from a large space of possible binary black hole populations.
Primary masses are assumed to follow a superposition between a power law and a Gaussian peak, secondary masses are power-law distributed, spin magnitudes and spin-orbit misalignment angles follow truncated Gaussians, and the merger rate is assumed to grow as a power law in $1+z$ (see Appendix~\ref{app:population-models} for these exact distributions and the range of hyperparameters chosen).
For each proposed population, we then compute the integrated detection efficiency $\xi(\Lambda)$ in two ways.
First, we estimate $\xi(\Lambda)$ via standard reweighting of the found binary black hole pipeline injections, as in Eq.~\eqref{eq:xi-reweighting} above.
Second, we instead compute the detection efficiency using our trained neural network, drawing an ensemble of binary parameters $\{\theta\} \sim \ppop$ from the proposed population and then directly evaluating the detection efficiency as in Eq.~\eqref{eq:mc-efficiency}.

Figure~\ref{fig:efficiency-check} shows the resulting detection efficiencies computed in both manners.
Each point corresponds to a randomly chosen population, and error bars correspond to expected Poisson uncertainties given the finite number of pipeline injections/draws used for each efficiency calculation.
In general, we see good agreement between $\xi(\Lambda)$ values obtained through traditional injection reweighting and values computed with our \pdettheta{} emulator across efficiencies spanning many orders of magnitude.
We do note that there are several points for which the two methods disagree, with the neural network predicting noticeably higher detection efficiencies than the reweighted pipeline injections.
Each of these points, though, has significant uncertainty in the reweighted pipeline injections' efficiency calculation, corresponding to populations that are very different from the reference distribution by which pipeline injections were drawn (specifically, these populations strongly favor unequal-mass binaries).

\section{Stabilizing hierarchical inference}
\label{sec:hierarchical-inference}

Hierarchical inference of the compact binary population can be limited by the accuracy with which the detection efficiency $\xi(\Lambda)$ can be estimated.
The detection efficiency is most commonly computed by reweighting a fixed set of pipeline injections, as in Eq.~\eqref{eq:xi-reweighting}.
Accurate estimation of $\xi(\Lambda)$ in this manner requires that (\textit{i}) the injections were drawn from a parent distribution, \pinj,that is ``close to'' the target population \ppop{}, and/or (\textit{ii}) that a very large number of found injections be available, which in turn requires a very large number of total trials $N_{\rm total}$. 

It is not clear what, formally, is meant by ``close to'' in the previous sentence.
What is clear, though, is that one or both of the above conditions can readily fail in practice, producing imprecise estimates of $\xi(\Lambda)$ and hampering inference of the compact binary population.
A common diagnostic is the ``effective number'' of injections informing an estimate of $\xi(\Lambda)$.
If we define $w_i = \ppopi/\pinji$ as a short-hand for the ratio appearing in Eq.~\eqref{eq:xi-reweighting}, then the number of effective injections is
    \begin{equation}
    N_{\rm eff} = \frac{\left(\sum_i w_i\right)^2}{\sum_j w_j^2},
    \label{eq:neff}
    \end{equation}
where both sums are again taken over found injections.
In order for systematic uncertainty in $\xi(\Lambda)$ due to a finite number of injections to remain a subdominant effect, it has been argued that one requires $N_{\rm eff} \gg c\,N_{\rm events}$, where $N_{\rm events}$ is the number of observed compact binaries and $c$ is some constant that is (hopefully) of order unity~\cite{Farr2019,Essick2022}.\footnote{In practice, a common choice is to demand that $N_{\rm eff} \geq 4 N_{\rm events}$, following Ref.~\cite{Farr2019}}
This implies that the total number of pipeline injections must scale linearly with catalog sizes, such that $N_{\rm total} \propto N_{\rm events}$.
However, other authors have argued that the number of injections must scale more steeply with catalog size, such that $N_{\rm total} \propto N_{\rm events}^\alpha$ with $1.5\lesssim \alpha \lesssim 2$~\cite{Talbot2023}.
Poorly-converged $\xi(\Lambda)$ estimates due to insufficient injections already and not infrequently limit our ability to explore the compact binary population.
The required increase of pipeline injections with catalog size (regardless of the precise scaling) implies that this issue will persist or be further exacerbated in the future.

\subsection{Dynamically drawing injections}
\label{sec:injection-regen-algorithm}

\begin{figure}
    \centering
    \includegraphics[width=0.48\textwidth]{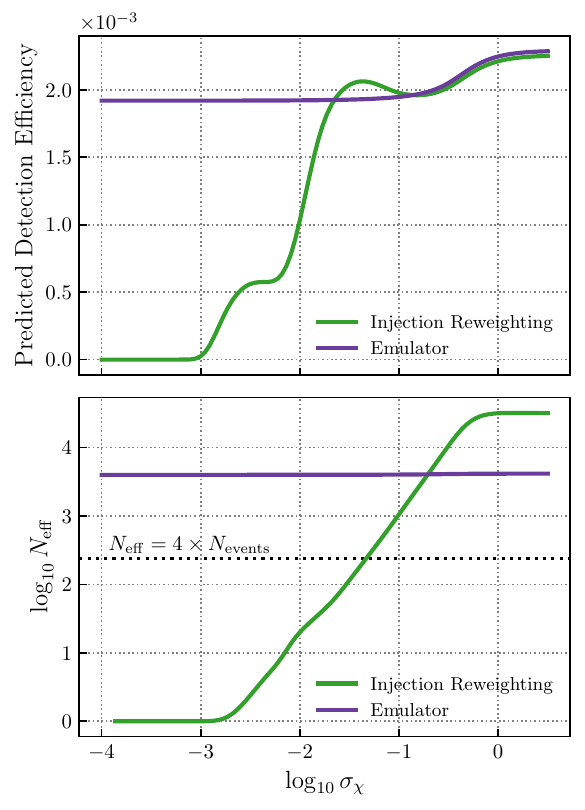}
    \caption{
    Illustration of convergence issues mitigated by use of a \pdettheta{} emulator in hierarchical inference.
    \textit{Top:} Predicted binary black hole detection efficiency $\xi$ as a function of the presumed width of the astrophysical spin magnitude distribution.
    Detection efficiencies are calculated via the reweighting of fixed pipeline injections (green) and by using the trained \pdettheta{} emulator to draw new injections at each value of $\sigma_\chi$ following the algorithm in Sec.~\ref{sec:injection-regen-algorithm} (purple).
    At large $\sigma_\chi$, both methods predict comparable detection efficiencies.
    As $\sigma_\chi$ decreases, however, the efficiencies predicted by injection reweighting drop unphysically to zero, due to the lack of injections falling inside the narrow range supported by the population model.
    \textit{Bottom:} The effective number of injections [see Eq.~\eqref{eq:neff}] informing each estimate.
    As $\sigma_\chi$ becomes small, the number of informative injections approaches zero, driving the unphysical behavior in the top panel.
    In particular, when $\sigma_\chi \lesssim 0.03$ we fail the convergence criteria $N_{\rm eff} \geq 4\times N_\mathrm{events}$ commonly adopted in the literature.
    When instead using the \pdettheta{} to draw new injections from each new distribution of interest, the number of effective injections remains approximately constant and the estimate of $\xi$ well-converged.
    }
    \label{fig:narrow-spins}
\end{figure}

A trained \pdettheta{} offers one avenue to mitigating the problem of poor $\xi(\Lambda)$ estimation.
The central problem, and the reason pipeline injections must grow in number with catalog size, is the fact that we typically must make do with a fixed injection set.
With a trained emulator, we can abandon this constraint and instead \textit{dynamically draw new injections} from each new population of interest.
A particularly efficient algorithm for doing this in the context of hierarchical inference is the following~\cite{Farr-private}:

\begin{enumerate}
    \item For each compact binary parameter, draw a large number of random values on the interval $[0,1]$:
        \begin{equation}
        \begin{aligned}
        &{\bm c}_{m_1} \sim U(0,1) \\
        &{\bm c}_{m_2} \sim U(0,1) \\
        &{\bm c}_z \sim U(0,1), \\
        &\mathrm{etc.}
        \end{aligned}
        \label{eq:cdf-sampling}
        \end{equation}
    where, e.g., ${\bm c}_{m_1} \equiv \{c_{m_1}\}$ indicates a vector of individual draws. 
    This is done once, prior to beginning inference.
    To improve convergence, these random values can be sampled jointly via low-discrepancy sequences, such as the Sobol sequence~\cite{NIEDERREITER198851}.
    \item Proceed with inference.
    Within the first likelihood evaluation with some proposed population $\Lambda$, compute, analytically or numerically, the inverse cumulative distribution function $F^{-1}_\Lambda(\cdot)$ associated with each compact binary parameter.
    \item Apply these inverse distributions to our draws from the unit interval to yield sets of physical parameter values;
                \begin{equation}
                \begin{aligned}
                {\bm m}_1 &= F^{-1}_{\Lambda,m_1}({\bm c}_{m_1}) \\
                {\bm m}_2 &= F^{-1}_{\Lambda,m_2}({\bm c}_{m_2}) \\
                {\bm z} &= F^{-1}_{\Lambda,z}({\bm c}_{z}).
                \end{aligned}
                \end{equation}
            The resulting values will be distributed according to the proposed population density, $p(\theta|\Lambda)$.
    \item Assemble these into a matrix ${\bm \Theta} = ({\bm m}_1\,\, {\bm m}_2\,\, {\bm z}\,\,...)$ of compact binary parameters with shape ($N_\mathrm{samp}, N_\mathrm{dim})$, and evaluate their detection probabilities with the trained neural network, yielding a vector of detection probabilities ${\bm P} = \hat{P}(\mathrm{det}|{\bm \Theta})$ with length $N_\mathrm{samp}$.
    \item Take the mean of the $N_\mathrm{samp}$ samples in ${\bm P}$ to obtain the detection efficiency: $\xi(\Lambda) = \langle {\bm P}\rangle $.
    \item Repeat Steps 2--5 for each subsequent likelihood evaluation.
\end{enumerate}

The scheme outlined above assumes a factorizable population model, such that the joint distribution $p(m_1, m_2, z, ...| \Lambda)$ can be written as the product $p(m_1|\Lambda) \,p(m_2|\Lambda)\, p(z|\Lambda)...\,$.
It can, however, be straightforwardly extended to non-factorizable populations with intrinsic correlations between parameters.
In this case, one iteratively performs inverse transform sampling using conditional probability distributions.
For example:
    \begin{enumerate}
    \item Compute the cumulative probability distribution $F_{\Lambda,z}(z)$ of source redshifts, and inverse transform sample to obtain a redshift $z$ drawn from $p(z|\Lambda)$.
    \item Given this redshift sample, define the conditional primary mass distribution $p(m_1|z,\Lambda)$.
    Compute the cumulative distribution of this conditional distribution, and inverse transform sample to obtain a primary mass drawn from $p(m_1|z,\Lambda)$.
    \item etc.
    \end{enumerate}
The result will be a tuple $\{z,m_1, m_2, ...\}$ drawn from the joint distribution $p(m_1,m_2,z,...|\Lambda)$.

Because the detection efficiency is evaluated using draws directly from the population $\Lambda$ of interest, the above algorithm can enable more accurate estimation of $\xi(\Lambda)$ than can be obtained through reweighting of fixed injections.
This is particularly true when attempting to investigate \textit{narrow} population features.
Narrow or abrupt features in the compact binary population are often of great astrophysical interest but are notoriously difficult to study, computationally speaking~\cite[e.g.][]{Callister2022}.
This is due, in part, to the fact that an estimate of $\xi(\Lambda)$ via reweighting of fixed injections will be necessarily be dominated by the small number of injections that happen to lie in the immediate vicinity of the feature of interest.
The resulting $\xi(\Lambda)$ will be subject to a small effective sample count and/or yield large log-likelihood variance.

\begin{figure*}
    \centering
    \includegraphics[width=0.8\textwidth]{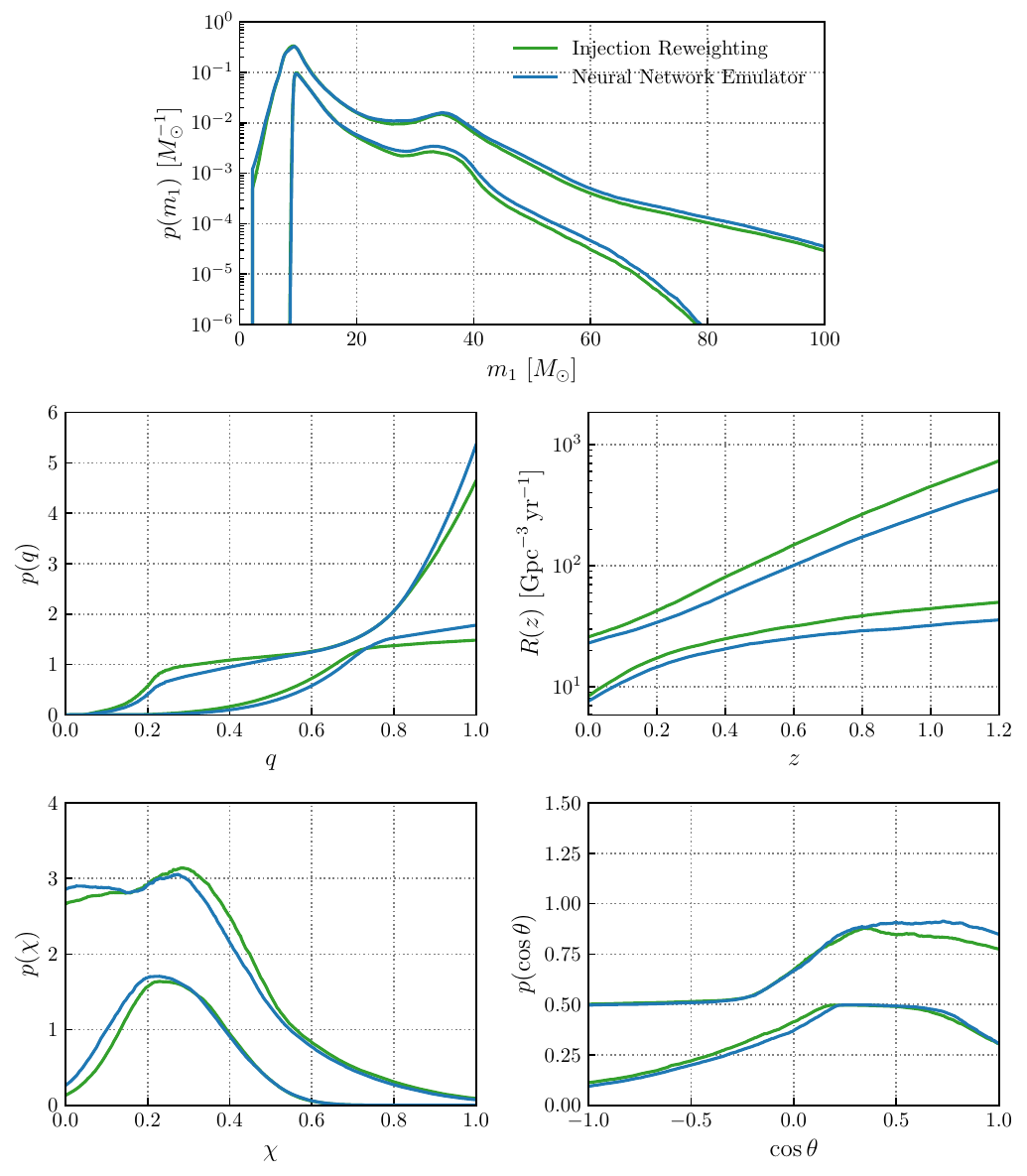}
    \caption{
    Measurements of the binary black hole population, using the 59 binary black holes detected in the LIGO-Virgo-KAGRA O3 observing run with false-alarm rates below $1\,\mathrm{yr}^{-1}$.
    Green curves show the central $95\%$ credible regions when correcting for selection effects via reweighting the fixed set of pipeline injections spanning the O3 observing run.
    Blue curves, meanwhile, show results obtained when instead using our trained \pdettheta{} emulator, along with the algorithm described in Sec.~\ref{sec:injection-regen-algorithm}, to dynamically draw \textit{new} found injections directly from the proposed population within each likelihood evaluation.
    The underlying posteriors on the hyperparameters governing the binary black hole mass, spin, and redshift distributions are shown in Appendix~\ref{app:more-results}; see Figs.~\ref{fig:corner-1}--\ref{fig:corner-3}.
    }
    \label{fig:inference-comparison}
\end{figure*}

The above algorithm, enabled by our \pdettheta{} emulator, avoids the poor convergence of $\xi(\Lambda)$ in the presence of narrow population features.
As a demonstration, the upper panel of Fig.~\ref{fig:narrow-spins} shows estimates of the detection efficiency for an observationally-plausible binary black hole population (see Appendix~\ref{app:population-models} for details) as we vary the assumed width $\sigma_\chi$ of the component spin magnitude distribution.
The purple curve shows the detection efficiency as calculated using our neural network emulator, following the algorithm above, while the green curve shows values obtained through reweighting of fixed pipeline injections.
The lower panel, meanwhile, shows the effective number of injections [Eq.~\eqref{eq:neff}] informing these estimates.

When $\sigma_\chi$ is large and the component spin distribution is broad, all is well:
both methods yield nearly equal $\xi(\Lambda)$ values and each is informed by a large number of effective samples, signifying that these values are robust.
As $\sigma_\chi$ is lowered and the spin distribution narrows, however, the injection reweighting begins to exhibit problems.
Within the lower panel, we see that the effective number of injections drops quickly; by the time we reach $\log_{10}\sigma \approx -1.5$, we are already falling below the $N_\mathrm{eff} \geq N_\mathrm{events}$ threshold often adopted for reliable inference.
For even smaller $\sigma_\chi$ we see the estimation of $\xi(\Lambda)$ break down.
The inferred detection efficiency rises briefly before plummeting unphysically to zero.
In contrast, direct evaluation with the neural network emulator remains well behaved, even as the component spin magnitude distribution approaches a delta function.

In principle, this approach is possible for semianalytic sensitivity estimates as well.
For example,~\citet{Essick2023} provides a closed-form estimate for $P(\mathrm{det}|\rho_\mathrm{opt})$ that accounts for the probability of different noise realizations.
However, in practice, this approach would require many waveform calls within each likelihood evaluation, which would be very costly.
The neural emulator avoids this by directly learning $P(\mathrm{det}|\theta)$ instead of $P(\mathrm{det}|\rho_\mathrm{opt})$.

Another advantage of the algorithm above is that it is \textit{differentiable}.
By drawing a fixed set of random values from the unit intervals (Step 1) and later inverse transforming sampling to obtain physical parameter values, we ensure that the likelihood is a deterministic function of $\Lambda$ and hence amenable to algorithms like Hamiltonian Monte Carlo.

We caution that the use of a detection probability emulator as described in this section is still subject to Monte Carlo variance.
Different realizations of random values in Eq.~\eqref{eq:cdf-sampling} will, in turn, yield slightly different estimated detection efficiencies.
This variance will decrease as one increases the number of Monte Carlo samples, but in some cases the required number of samples may be large, particularly if binary detections come primarily from a very small (and hence improbable) portion of parameter space.
For example, the integrated detection efficiency is usually dominated by the small fraction of events at low redshift, whereas the vast majority of events under reasonable population models occur at high redshifts; a very large number of samples will therefore be needed to obtain a reasonable number of nearby events.
In such cases, one can instead adopt a variant of the algorithm described above, in which some parameters are inverse-transform sampled directly from the proposed population $p(\theta|\Lambda)$ while others (like redshift) are reweighted from a fixed reference distribution.
Such a hybrid strategy can improve convergence and decrease the overall number of samples required to estimate $\xi(\Lambda)$.
More details are provided in Appendix~\ref{app:hierarchical-inference}.

\subsection{Full hierarchical inference: A demonstration}

As a further demonstration of this approach, as well as a test of our \pdettheta{} emulator, we perform complete hierarchical inference of the binary black hole population using the algorithm described above to compute $\xi(\Lambda)$.
We use the 59 binary black holes observed during the LIGO-Virgo O3 observing run with false-alarm rates below $1\,\mathrm{yr}^{-1}$~\cite{GWTC3,abbott_open_2023},\footnote{The events GW190814~\cite{GW190814} and GW190917~\cite{GWTC2-1}, with their very uneven mass ratios and low secondary masses, are excluded as outliers relative to the bulk binary black hole population~\cite{O3b-pop}.} and adopt population models comparable to those used in recent LIGO-Virgo-KAGRA Collaboration analyses~\cite{O3b-pop}.
Specifically, we assume that source-frame primary masses follow a mixture between a power-law continuum and a Gaussian peak~\cite{Talbot2018} and that secondary masses are power-law distributed~\cite{Fishbach2020}.
Component spin magnitudes follow a truncated Gaussian distribution, while cosine spin-orbit tilts are described as a mixture between an isotropic component and a Gaussian excess~\cite{O2-pop}.
The source-frame binary black hole volumetric merger rate is assumed to follow a power law in $(1+z)$~\cite{Fishbach2018}.
The exact population models used and the priors on their parameters are presented in Appendix~\ref{app:population-models}.

Results are shown in Fig.~\ref{fig:inference-comparison}.
The pair of green curves shows the $95\%$ credible constraints on the probability distribution/merger rates of binary black holes using standard injection reweighting.
Blue curves show constraints instead obtained using the trained \pdettheta{} emulator.
Both sets of results are near matches, with the emulator yielding accurate reconstruction of the binary black hole primary mass, mass ratio, spin magnitude, and spin tilt distributions, as well as accurate reconstruction of the redshift-dependent merger rate.
Posteriors on the underlying hyperparameters under both approaches can be seen in Appendix~\ref{app:hierarchical-inference}.

While the results in Fig.~\ref{fig:inference-comparison} are consistent with one another, they are not exact matches.
In particular, the neural network-based selection effects yield a slightly stronger preference for equal mass ratios and slightly less evolution of the merger rate with redshift, relative to selection effects estimated via injection reweighting.
It is not clear which approach is more accurate.
On the one hand, the neural network emulator may be more reliably interpolating the underlying selection function, particularly in regions where injections are sparse (such as low mass ratios).
On the other hand, because the neural network was trained on the same injections informing the results in green, we might expect a perfectly performing emulator to yield identical results (up to variance associated with Monte Carlo averaging).
The slight differences in Fig.~\ref{fig:inference-comparison} may therefore indicate further room for improvement.

\section{Comparison to Semianalytic Selection Effects}
\label{sec:semianalytic}

\begin{figure*}
    \centering
    \includegraphics[width=\textwidth]{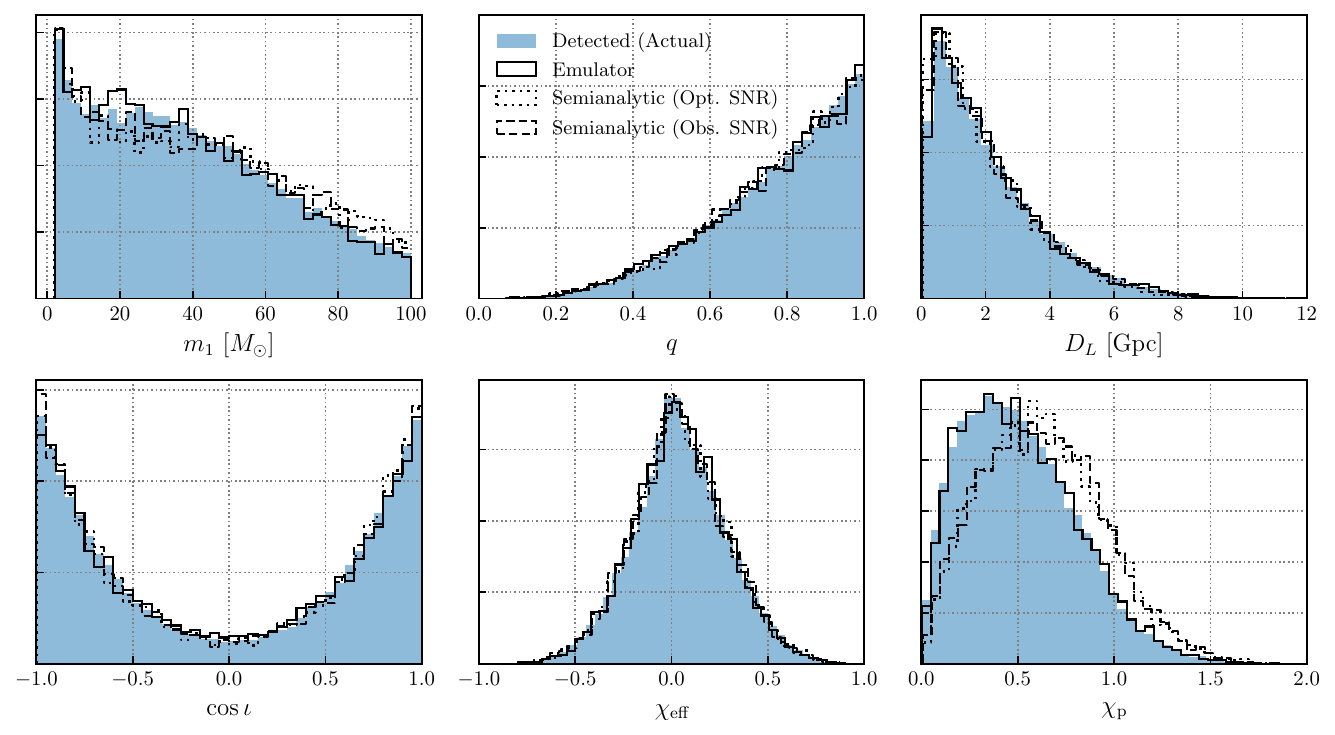}
    \caption{
    As in Fig.~\ref{fig:bbh}, but additionally detected binary black hole properties predicted by approximating Advanced LIGO \& Advanced Virgo selection effects via a threshold on matched filter signal-to-noise ratio.
    Dotted distributions show results when demanding an optimal SNR $\rho_{\rm opt} \geq 10$, while dashed distributions correspond to a threshold $\hat \rho \geq 10$ on randomly-perturbed ``observed'' SNRs that capture the effects of fluctuating noise~\cite{Essick2023}.
    With the exception of the effective precessing spin parameter, semianalytic SNR thresholds can broadly capture the distributions of detectable binary black holes, with visual matches comparable to our trained neural network emulator.
    However, we see that SNR thresholds systematically underestimate the sensitivity of the LIGO-Virgo network to low-mass binaries at large distances, whereas the neural network does not; see Fig.~\ref{fig:dist-mass}.
    }
    \label{fig:bbh-semianalytic}
\end{figure*}

\begin{figure*}
    \centering
    \includegraphics[width=\textwidth]{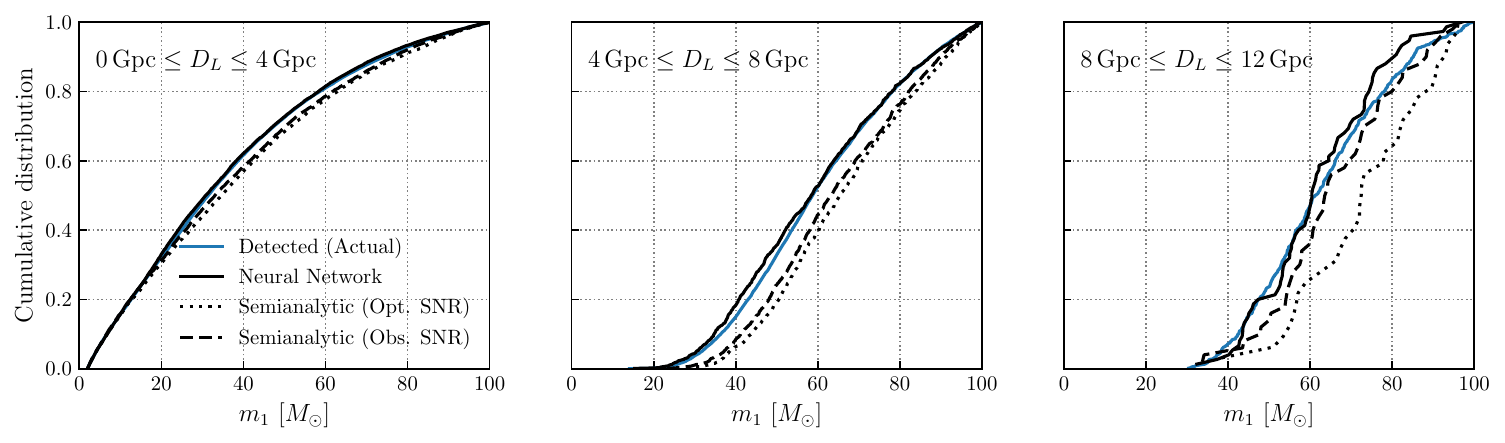}
    \caption{
    Cumulative distributions of detected binary black hole primary masses in three different luminosity distance intervals, as predicted by real pipeline injections (blue), the trained $\pdettheta{}$ emulator (solid black), and semianalytic thresholds on optimal and observed SNRs (dotted and dashed black, respectively).
    In all distance intervals, semianalytic sensitivity estimates underestimate the detectability of low-mass binaries, shifting predicted cumulative distributions to the right.
    }
    \label{fig:dist-mass}
\end{figure*}

As noted in Sec.~\ref{sec:intro}, it is common for gravitational-wave selection effects to be semianalytically approximated via a threshold on a source's matched filter SNR.
This threshold is often placed on a source's optimal SNR or on a simulated realization of an ``observed'' SNR that mimics random fluctuations due to noise.
Although SNR thresholds can approximate search selection effects, they are known to neglect higher order aspects of gravitational-wave detection, including signal consistency checks and non-stationary noise.
In this section, we compare our trained $\pdettheta{}$ emulator to traditional semianalytic SNR thresholds, exploring the degree to which the neural network learns additional, higher order information not contained in SNRs alone.

In Figure~\ref{fig:bbh-semianalytic}, we again show the distribution of successfully detected binary black hole injections (blue) together with predictions from our trained neural network emulator (solid black); see Fig.~\ref{fig:bbh}.
Each panel of this figure contains two additional curves.
The dotted histograms show properties of found events as predicted by a semianalytic threshold on source's optimal SNRs.
Specifically, an ensemble of simulated events is drawn from the same parent distribution as the real pipeline injections.
Each simulated event is placed at a random time during the O3 observing run, and its optimal matched filter SNR is computed, summing in quadrature over the three Advanced LIGO and Advanced Virgo detectors.
Sources are labeled as ``detected'' if their optimal SNRs exceed $\rho_{\rm opt}\geq 10$, a threshold found to broadly approximate the $1\,\mathrm{yr}^{-1}$ false-alarm rate threshold adopted in this and many other works.
The dashed histograms are analogous, but constructed by instead demanding that simulated ``observed'' network SNRs exceed $\hat \rho\geq 10$, following Ref.~\cite{Essick2023}.
These observed SNRs are randomly drawn from the probability distribution of possible SNRs for each event, accounting for the effects of random noise fluctuations.\footnote{The quantity $\hat \rho$ is referred to as $\rho_{\mathrm{net},\phi}$ in Ref.~\cite{Essick2023}.}

Within Fig.~\ref{fig:bbh-semianalytic}, we see that, while our trained $\pdettheta{}$ emulator yields the correct distribution of effective precessing spin parameters, the semianalytic SNR thresholds do not.
This may reflect the fact that matched filtering template banks do not generally include effects of spin precession, and thus semianalytic calculations may systematically overestimate the SNRs of strongly-precessing binaries.
Barring $\chi_{\rm p}$, however, it appears that the semianalytic thresholds on both optimal and observed SNRs broadly recover realistic distributions of detected binary black holes, performing comparably to our trained emulator.

This conclusion breaks down, however, if we more carefully consider predicted detections as a function of distance.
Figure~\ref{fig:dist-mass} shows, in blue, the cumulative probability distribution of found pipeline injections within three consecutive luminosity distance shells.
Also shown are predictions from the neural network emulator (solid black), the semianalytic optimal SNR cut (dotted black), and the semianalytic observed SNR cut (dashed black).
As we move to larger distances, we see that a semianalytic threshold on optimal SNRs predicts binary black hole detections systematically shifted towards larger masses.
In other words, this strategy \textit{systematically underestimates} the sensitivity of Advanced LIGO \& Advanced Virgo to low-mass binaries and \textit{systematically overestimates} the sensitivity to high-mass binaries.
This bias is lessened by instead thresholding on simulated observed SNRs, but it remains present.
This behavior is consistent with Ref.~\cite{Essick2023}, which found that a \textit{mass-dependent} SNR threshold was needed to accurately predict the distance distribution of detected binary black holes, with higher-mass binaries requiring higher semianalytic thresholds (see their Fig.~8).

The trained neural network, in contrast, produces accurate cumulative distributions in all distance bins (the elevated variance in the $8$--$12\,\mathrm{Gpc}$ interval is due to a very small number of events in this range).
We therefore conclude that the $\pdettheta{}$ emulator is successfully learning higher-order information encoded in pipeline injections but not captured by a simple SNR threshold.

\begin{figure*}
    \centering
    \includegraphics[width=\textwidth]{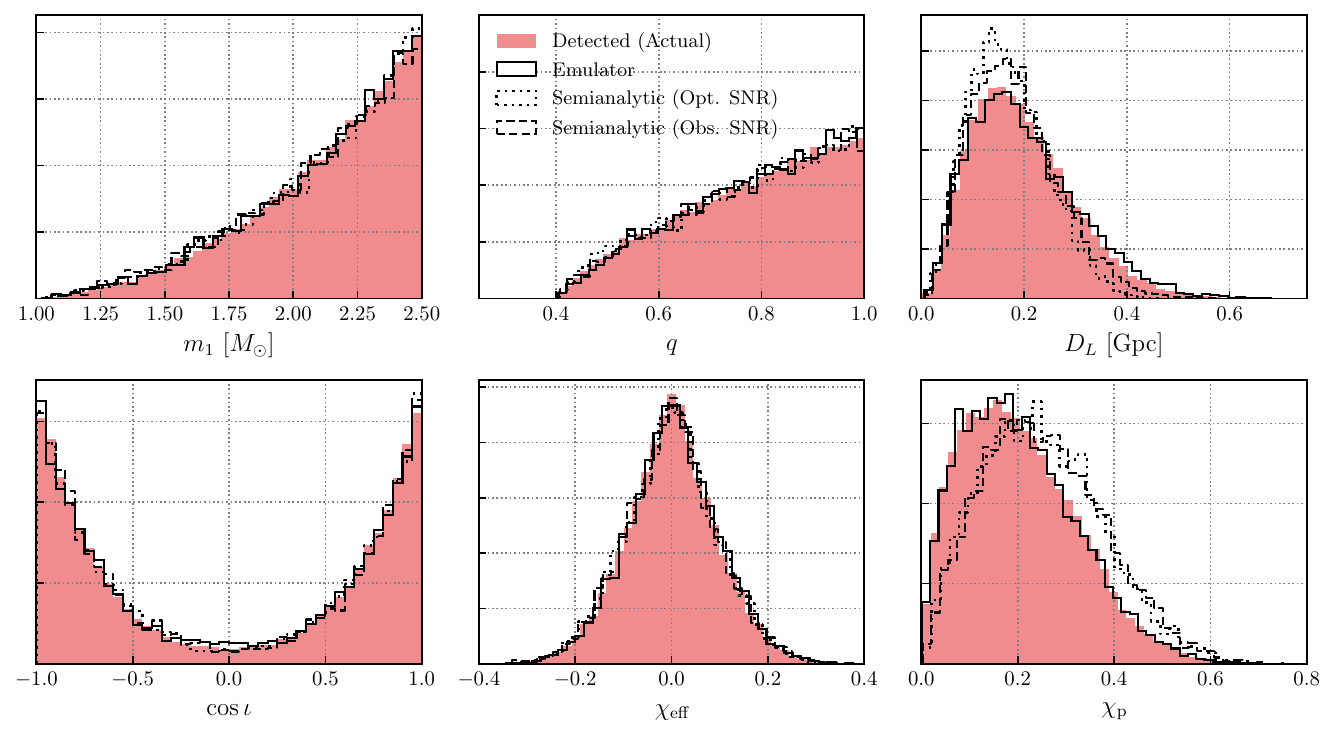}
    \caption{
    As in Fig.~\ref{fig:bbh-semianalytic}, but for binary neutron stars.
    }
    \label{fig:bns-semianalytic}
\end{figure*}

\begin{figure*}
    \centering
    \includegraphics[width=\textwidth]{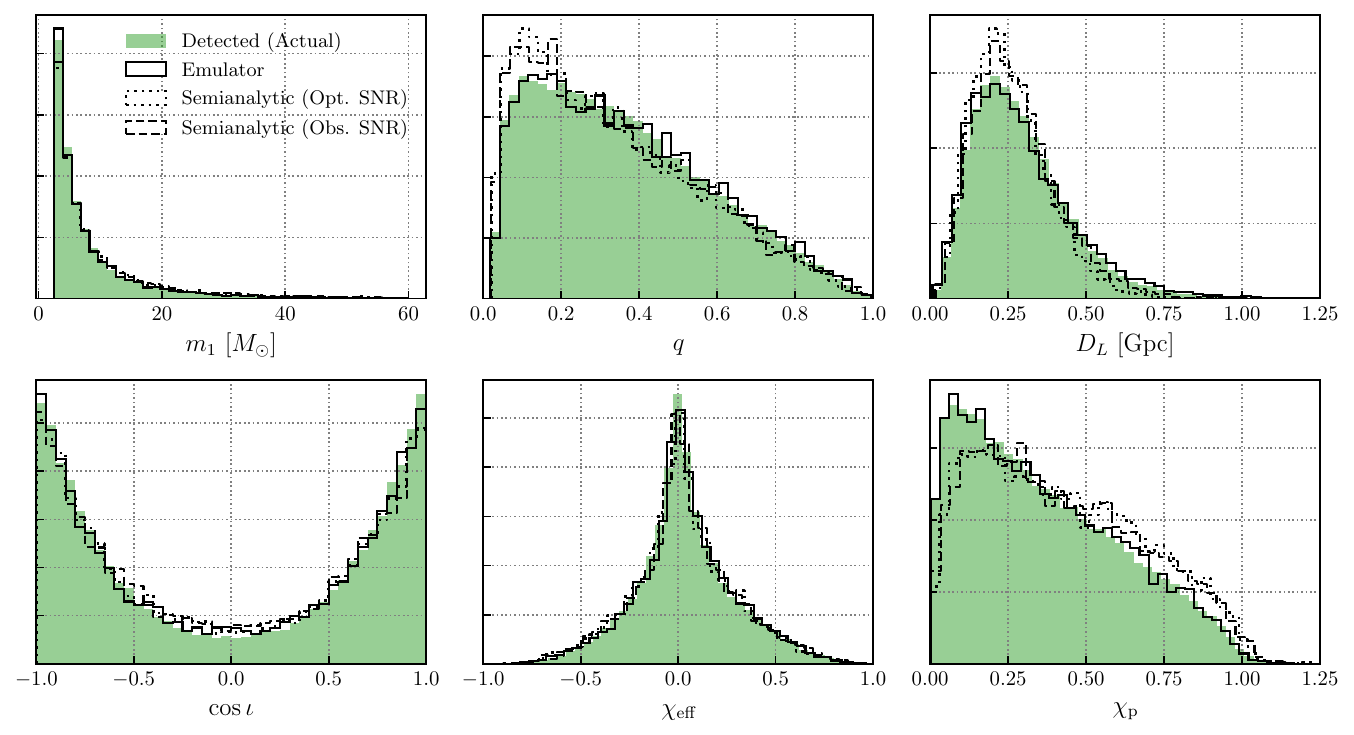}
    \caption{
    As in Fig.~\ref{fig:bbh-semianalytic}, but for neutron star-black hole binaries.
    }
    \label{fig:nsbh-semianalytic}
\end{figure*}

We show analogous results for binary neutron stars and neutron star-black hole binaries in Figs.~\ref{fig:bns-semianalytic} and \ref{fig:nsbh-semianalytic}, respectively.
The trends identified above for binary black holes persist: semianalytic SNR thresholds tend to poorly predict effective precessing spin distributions and underpredict the distances to which low-mass systems are successfully detected, while the neural network emulator more accurately matches found pipeline injections.
Within Fig.~\ref{fig:nsbh-semianalytic}, we also see that semianalytic approximations overpredict the sensitivity of Advanced LIGO \& Advanced Virgo to systems with very unequal masses, sensitivity more accurately captured by the $\pdettheta{}$ emulator.

We note that, although (possibly parameter-dependent) semianalytic SNR thresholds and our neural network emulator may successfully predict similar distributions of compact binary detections, SNR thresholds \textit{cannot} compare with the neural network emulator in hierarchical inference.
As described in Sec.~\ref{sec:hierarchical-inference}, the utility of the neural network is its ability to estimate $P(\mathrm{det}|\theta)$ for a new ensemble of binaries in each new likelihood evaluation.
This, in turn, requires a $P(\mathrm{det}|\theta)$ that is computationally efficient and, in modern computing environments, differentiable.
Detection probability estimation via SNR thresholding satisfies neither of these requirements.
In particular, the calculation of signal-to-noise ratios requires the evaluation of gravitational waveforms; waveform generation is usually slow and non-differentiable, and thus cannot be used in an algorithm like that described in Sec.~\ref{sec:injection-regen-algorithm}.

\section{Discussion}
\label{sec:discussion}

In this paper we have presented a neural network emulator with which to describe the detection probabilities of compact binary mergers in the LIGO-Virgo-KAGRA O3 Observing Run.
We described the construction and training of the emulator, and validated its accuracy via complete hierarchical inference of the binary black hole population.

This is not the first such tool in the literature;
a number of previous studies have pursued various machine learning strategies to model gravitational-wave selection effects~\cite{Gerosa2020,Talbot2022,Wong2021,Mould2022,ChapmanBird2023}.
Reference~\cite{Gerosa2020} trained a neural network to classify compact binaries as detectable or undetectable.
Like our study, this classification was performed on a per event basis.
One limitation of that work, however, was its reliance on an idealized detection process: Binaries were regarded as ``detectable'' if they exceeded an optimal matched filtering signal-to-noise ratio threshold, calculated using fixed instrumental noise power spectral densities.
In reality, drifting instrumental sensitivities, terrestrial noise transients, and additional signal consistency requirements mean that compact binary detection is more complicated than a universal signal-to-noise ratio threshold.
While the emulators presented in Ref.~\cite{Gerosa2020} are likely sufficient for qualitatively accurate forecasting of future detections, they likely do not capture these higher order effects required for precise inference of the binary population; for this, injection campaigns (like those comprising our training data) are needed.

A complementary strategy was undertaken in Ref.~\cite{Talbot2022}, in which \pdettheta{} emulation is framed as a problem of \textit{density estimation}.
In this approach, a Gaussian mixture model was fit to the distribution of found injections, yielding a function proportional to the product $\pdettheta{} \pinj$, where \pinj{} is the distribution from which the injections were drawn.
Relative detection probabilities of \textit{new} injections $\theta'$ could then be calculated by evaluating the mixture model and dividing by $p(\theta'|\Lambda_\mathrm{inj})$, leaving values proportional to $P(\mathrm{det}|\theta')$.
This density estimation approach has the advantage of being computationally efficient and inexpensive to train.
One disadvantage is the requirement that users must track and evaluate the draw probabilities $p(\theta'|\Lambda_\mathrm{inj})$.
This becomes difficult when training injections are themselves complex mixtures between many disparate sets of events (as in Fig.~\ref{fig:training}).
By learning the latent \pdettheta{} surface itself, we avoid the downstream need to reevaluate $\pinj{}$.
Directly learning \pdettheta{} correspondingly facilitates iterative learning:
if the emulator is found to perform poorly in a specific neighborhood of parameter space, additional training data can be generated in that neighborhood and the emulator retrained.
This cycle can be repeated as necessary, with no ties to a globally defined \pinj{} distribution.

Instead of learning \pdettheta{}, Refs.~\cite{Wong2021,Mould2022} directly emulated the integrated detection efficiency $\xi(\Lambda)$ as a function of chosen hyperparameters $\Lambda$.
Reference~\cite{ChapmanBird2023} combined this idea with $\pdettheta{}$ emulation in a two step process, training a neural network to emulate signal-to-noise ratios that were then used as training inputs for a second network emulating $\xi(\Lambda)$.
The direct emulation of detection efficiencies sidesteps the (computationally burdensome) need to generate any injections and may help to avoid convergence issues associated with Monte Carlo integration.
At the same time, the direct emulation of $\xi(\Lambda)$ commits oneself to a single, chosen model for the compact binary population, characterized by a specific set of hyperparameters $\Lambda$.
Inference of the compact binary population with a new model would require retraining of the detection efficiency emulator, a process we wish to avoid.

As discussed above, an altogether different approach that does not rely on machine learning is to approximate search selection via a signal-to-noise ratio threshold~\cite{Essick2023,Gerosa2024}.
As we demonstrated in Sec.~\ref{sec:semianalytic}, a uniform SNR threshold applied across binary parameter space does not accurately reproduce distributions of found binary parameters, and will therefore introduce biases if used as a proxy for selection effects when analyzing gravitational-wave data.
However, good performance might be achievable by calibrating a source-dependent threshold that varies across the space of binary parameters~\cite{Essick2023}.
Alternatively, there exist analytic fitting functions that capture the $\pdettheta{}$ on the lower-dimensional space of binary masses, distance, and aligned spin components~\cite{LorenzoMedina2024}.
Analytic methods like these have the advantage of being directly interpretable.
A deep learning approach, on the other hand, trades some interpretability for speed, flexibility, and ease of generalization: a neural network like the one we present here can capture the behavior of $\pdettheta{}$ across the high-dimensional space of compact binary parameters (including spin precession and extrinsic parameters) while requiring no direct waveform evaluation.

As continuous representations of \pdettheta{} grow in prevalence, they may influence the metrics by which pipeline injection sets are designed.
Current injection sets are carefully designed to maximize the efficiencies with which they can be reweighted to other populations of interest, as in Eq.~\eqref{eq:xi-reweighting}.
When using pipeline injections as training data for \pdettheta{} representations (whether a signal-to-noise threshold, an analytic fitting function, or a neural network emulator), though, the best performance may be achieved with alternative design metrics.
In the case of a neural network emulator, for example, it is beneficial to have injections uniformly placed across a much broader range of parameter space and include both very loud and very quiet events.
It would be valuable for future studies to more quantitatively explore suitable metrics for the generation of training data.

Similarly, future studies should explore how emulator accuracy scales with the number of pipeline injections provided as training data.
In the Advanced LIGO and Advanced Virgo O3 observing run, $2.5\times10^5$ pipeline injections were performed for each of the binary black hole, binary neutron star, and neutron star-black hole populations.
We used just under half of these when training our emulator (see Table~\ref{tab:inj-population}).
If emulator precision could be maintained while further decreasing the number of training injections, this may minimize the future computational cost of calibrating gravitational-wave selection effects.
It will also be valuable to more systematically explore different network architectures.
In our study, we heuristically found that a fully-connected network with four 192-neuron wide hidden layers yielded a reasonable balance between predictive accuracy and training time.
It is possible, though, that further advances (e.g. an improved loss function, alternate training data, etc.) could enable comparable accuracy with a smaller network.

Our trained detection probability emulator is made publicly available at \url{https://github.com/tcallister/pdet}~\cite{pdet}.
The use of this emulator is described in Appendix~\ref{sec:instructions} below, with more details found in documentation online.

\begin{acknowledgments}
We thank Muhammad Zeeshan, Ben Farr, Christopher Berry, and Matthew Mould for valuable comments and conversation.
T.~C.~is supported by the Eric and Wendy Schmidt AI in Science Postdoctoral Fellowship, a Schmidt Sciences program.
R.~E.~is supported by the Natural Sciences \& Engineering Research Council of Canada (NSERC) through a Discovery Grant (RGPIN-2023-03346).
D.~E.~H. is supported by NSF grants AST-2006645 and PHY2110507, as well as by the Kavli Institute for Cosmological Physics through an endowment from the Kavli Foundation and its founder Fred Kavli.
This work was completed in part with resources provided by the University of Chicago’s Research Computing Center.
The authors are also grateful for additional computational resources provided by the LIGO Laboratory and supported by National Science Foundation Grants PHY-0757058 and PHY-0823459.
This material is based upon work supported by NSF's LIGO Laboratory which is a major facility fully funded by the National Science Foundation.\\

\textit{Data \& code availability:}
Our trained detection probability emulator is available at \url{https://github.com/tcallister/pdet}~\cite{pdet}, and code used to produce the results in this study can be found at \url{https://github.com/tcallister/learning-p-det}~\cite{learning-pdet}.
The necessary data to regenerate figures or rerun analyses are available via Zenodo~\cite{data-release}.
\\

\textit{Software used:}
This work made use of the following software packages:
\texttt{arviz}~\citep{arviz_2019},
\texttt{astropy}~\citep{astropy:2013, astropy:2018, astropy:2022},
\texttt{cython}~\citep{cython:2011},
\texttt{h5py}~\citep{collette_python_hdf5_2014, h5py_7560547},
\texttt{jax}~\citep{jax},
\texttt{matplotlib}~\citep{Hunter:2007},
\texttt{numpy}~\citep{numpy},
\texttt{numpyro}~\citep{numpyro1,numpyro2},
\texttt{pandas}~\citep{mckinney-proc-scipy-2010, pandas_10957263}, 
\texttt{python}~\citep{python}, 
\texttt{scikit-learn}~\citep{scikit-learn, sklearn_api, scikit-learn_11237090},
\texttt{scipy}~\citep{2020SciPy-NMeth, scipy_11255513},
and \texttt{tensorflow}~\citep{tensorflow_12119782}.
Software citation information aggregated using \texttt{\href{https://www.tomwagg.com/software-citation-station/}{The Software Citation Station}} \citep{software-citation-station-paper, software-citation-station-zenodo}.

\end{acknowledgments}

\appendix

\section{Using the emulator}
\label{sec:instructions}

A \texttt{python} implementation of the trained detection probability emulator is available at \url{https://github.com/tcallister/pdet}~\cite{pdet}.
In this section, we briefly describe how to access and use this the trained network.

Most directly, the trained emulator can be directly imported and evaluated as illustrated in the following example:

\begin{minted}[mathescape,
               gobble=0,
               frame=lines,
               framesep=3mm]{python}
from p_det import p_det_O3

# Instantiate trained emulator
p = p_det_O3()

# Define data.
# The following shows a minimal working example,
# in which we specify source-frame component
# masses, spin magnitudes, and redshifts for
# three compact binaries
params = {'mass_1':[2.5,10.0,15.0],
          'mass_2':[1.2,5.0,10.0],
          'a_1':[0.0,0.2,0.3],
          'a_2':[0.1,0.4,0.2],
          'redshift':[0.1,0.9,1.0]
          }

# Compute detection probabilities
detection_probs = p.predict(params)
\end{minted}

In this example, the user has provided the minimum set of required parameters: (\textit{i}) source-frame component masses, (\textit{ii}) component spin magnitudes, and (\textit{iii}) a distance parameter (either redshift, luminosity distance, or comoving distance).
Additional quantities like spin orientations and extrinsic parameters can be optionally provided; if not provided they are randomly generated assuming isotropy.
In this example, compact binary parameters were provided in the form of a dictionary, but they may also be passed via any other structure supporting key-value functionality. 
Internally, the \texttt{p\_det\_O3.predict} method checks for the presence and self-consistency of provided parameters, transforms to the input parameter space expected by the neural network, and evaluates the network.

The above example illustrates how one might use the trained network when forward modeling sets of observable compact binary signals.
As in Sec.~\ref{sec:hierarchical-inference}, another use case is to employ the network in hierarchical inference of the compact binary population.
To this end, we need an interface that is, ideally, compileable and differentiable, to enable compatibility with model likelihoods and inference performed with \texttt{jax}~\cite{jax} and \texttt{numpyro}~\cite{numpyro1,numpyro2}.
This is provided by calling \texttt{p\_det\_O3} directly (which implicitly evaluates the \texttt{p\_det\_O3.\_\_call\_\_} method) as follows:
\begin{minted}[mathescape,
               gobble=0,
               frame=lines,
               framesep=3mm]{python}
from p_det import p_det_O3
import jax
import jax.numpy as jnp

# Instantiate trained emulator
p = p_det_O3()

# Obtain just-in-time-compiled probability
jitted_p_det_O3 = jax.jit(p)

# Generate and define binary parameters.
# See online documentation for the proper
# contents and formatting of this object
mass_1 = [20., 30., ...]
mass_2 = [15., 29., ...]
a_1 = [0.5, 0.9, ...]
a_2 = [0.3, 0., ...]
...
params = jnp.array([mass_1,
                    mass_2,
                    a_1,
                    a_2,
                    ...])

# Compute detection probabilities
detection_probs = jitted_p_det_O3(params)
\end{minted}
Direct evaluation in this manner necessarily lacks the guardrails and self-consistency checks built into the \texttt{p\_det\_O3.predict} method, and instead assumes that users follow a specific, expected format in providing compact binary parameters; see code documentation for exact details.

\section{More On Emulator Training}
\label{app:tuning}

This appendix provides additional details regarding the training data, loss function, and procedure used for neural network training.

\subsection{Training data}
\label{app:training-data}

\begin{table*}[]
    \setlength{\tabcolsep}{4pt}
    \renewcommand{\arraystretch}{1.2}
    \centering
    \footnotesize
    \caption{
    Description of injection sets used to train \pdettheta{} emulator, including the hyperparameters defining each set, the convention followed when defining component mass distributions, and the number $N$ of injections used from each set; see Sec.~\ref{app:training-data}.
    Note that, because $m_{2,{\rm max}} = m_{1,{\rm min}}$ for the neutron star-black hole injections, conventions~\eqref{eq:p-m2-conditional} and \eqref{eq:p-m2-not} are equivalent for this population.
    }
    \begin{tabular}{r | l l l l l l l l l l l l | l}
    \hline \hline
    Injection Set 
        & $m_{1,{\rm min}}$
        & $m_{1,{\rm max}}$ 
        & $\alpha$
        & $m_{2,{\rm min}}$
        & $m_{2,{\rm max}}$ 
        & $\beta_q$ 
        & Convention
        & $\chi_{1,{\rm max}}$ 
        & $\chi_{2,{\rm max}}$
        & $\kappa$ 
        & $z_{\rm max}$
        & $f_{\rm ref}$
        & $N$ \\
    \hline
    Pipeline BBH
        & $2\,M_\odot$
        & $100\,M_\odot$ 
        & $-2.35$
        & $2\,M_\odot$
        & $100\,M_\odot$ 
        & $1$ 
        & Eq.~\eqref{eq:p-m2-conditional}
        & $0.998$ 
        & $0.998$ 
        & $1$ 
        & $1.9$
        & $10\,\mathrm{Hz}$
        & $9\times10^4$\\
    Hopeless BBH
        & $2\,M_\odot$
        & $100\,M_\odot$ 
        & $-1$
        & $2\,M_\odot$
        & $100\,M_\odot$ 
        & $1$ 
        & Eq.~\eqref{eq:p-m2-conditional}
        & $0.998$ 
        & $0.998$ 
        & $0$ 
        & $1.9$
        & $16\,\mathrm{Hz}$
        & $1.4\times10^5$ \\
    Certain BBH
        & $2\,M_\odot$
        & $100\,M_\odot$ 
        & $-1$
        & $2\,M_\odot$
        & $100\,M_\odot$ 
        & $1$ 
        & Eq.~\eqref{eq:p-m2-conditional}
        & $0.998$ 
        & $0.998$ 
        & $0$ 
        & $1.9$
        & $16\,\mathrm{Hz}$
        & $1.4\times10^5$ \\
    \hline
    Pipeline NSBH
        & $2.5\,M_\odot$
        & $60\,M_\odot$ 
        & $-2.35$
        & $1\,M_\odot$
        & $2.5\,M_\odot$ 
        & $0$ 
        & Eq.~\eqref{eq:p-m2-conditional}
        & $0.998$ 
        & $0.4$ 
        & $0$ 
        & $0.25$
        & $15\,\mathrm{Hz}$
        & $9\times10^4$ \\
    Hopeless NSBH
        & $2.5\,M_\odot$
        & $60\,M_\odot$ 
        & $-1$
        & $1\,M_\odot$
        & $2.5\,M_\odot$ 
        & $0$ 
        & Eq.~\eqref{eq:p-m2-conditional}
        & $0.998$ 
        & $0.4$ 
        & $0$ 
        & $0.25$
        & $16\,\mathrm{Hz}$
        & $1.4\times10^5$ \\
    Certain NSBH
        & $2.5\,M_\odot$
        & $60\,M_\odot$ 
        & $-1$
        & $1\,M_\odot$
        & $2.5\,M_\odot$ 
        & $0$ 
        & Eq.~\eqref{eq:p-m2-conditional}
        & $0.998$ 
        & $0.4$ 
        & $0$ 
        & $0.25$
        & $16\,\mathrm{Hz}$
        & $1.4\times10^5$ \\
    \hline
    Pipeline BNS
        & $1\,M_\odot$
        & $2.5\,M_\odot$ 
        & $0$
        & $1\,M_\odot$
        & $2.5\,M_\odot$ 
        & $0$ 
        & Eq.~\eqref{eq:p-m2-not}
        & $0.4$ 
        & $0.4$ 
        & $0$ 
        & $0.15$
        & $15\,\mathrm{Hz}$
        & $9\times10^4$ \\
    Hopeless BNS
        & $1\,M_\odot$
        & $2.5\,M_\odot$ 
        & $0$
        & $1\,M_\odot$
        & $2.5\,M_\odot$ 
        & $0$ 
        & Eq.~\eqref{eq:p-m2-not}
        & $0.4$ 
        & $0.4$ 
        & $0$ 
        & $0.15$
        & $16\,\mathrm{Hz}$
        & $1.4\times10^5$ \\
    Certain BNS
        & $1\,M_\odot$
        & $2.5\,M_\odot$ 
        & $0$
        & $1\,M_\odot$
        & $2.5\,M_\odot$ 
        & $0$ 
        & Eq.~\eqref{eq:p-m2-not}
        & $0.4$ 
        & $0.4$ 
        & $0$ 
        & $0.15$
        & $16\,\mathrm{Hz}$
        & $1.4\times10^5$ \\
    \hline
    Auxiliary Hopeless
        & $1\,M_\odot$
        & $100\,M_\odot$ 
        & $-2$
        & $1\,M_\odot$
        & $100\,M_\odot$ 
        & $-2$ 
        & Eq.~\eqref{eq:p-m2-not}
        & $0.998$ 
        & $0.998$ 
        & $-1$ 
        & $2$
        & $16\,\mathrm{Hz}$
        & $2.4\times10^5$ \\
    \hline
    \hline
    \end{tabular}
    \label{tab:inj-population}
\end{table*}

As discussed in the main text, our training data comprises sets of simulated compact binaries added to Advanced LIGO and Advanced Virgo data, analyzed with search pipelines, and labeled as detected (found) or undetected (missed); see Fig.~\ref{fig:cartoon}.
Briefly, each injected population is described via a power-law primary mass distribution,
    \begin{equation}
    p(m_1|\Lambda_\mathrm{inj}) \propto m_1^\alpha \qquad (m_{1,{\rm min}} \leq m_1 \leq m_{1,{\rm max}}).
    \label{eq:p-m1 powerlaw}
    \end{equation}
Secondary masses are also described as a power laws, following one of two conventions.
First, the secondary mass distribution can be defined conditionally on $m_1$, such that
    \begin{equation}
    p(m_2|m_1, \Lambda_\mathrm{inj}) \propto m_2^{\beta_q} \qquad (m_{2,{\rm min}} \leq m_2 \leq m_1),
    \label{eq:p-m2-conditional}
    \end{equation}
Alternatively, we can directly describe the joint distribution of $m_1$ and $m_2$ as
    \begin{equation}
    p(m_1,m_2|\Lambda_\mathrm{inj}) \propto m_1^\alpha m_2^{\beta_q} \Theta(m_1 - m_2).
    \label{eq:p-m2-not}
    \end{equation}
Here, $\Theta(\cdot)$ is the Heaviside step function.
Note, Eq.~\ref{eq:p-m2-not} is a different distribution than the product of Eqs.~\ref{eq:p-m1 powerlaw} and~\ref{eq:p-m2-conditional}.
All injection sets have independently and identically-distributed component spins, with isotropic spin orientations and uniform spin magnitude distributions between $0\leq \chi_1 \leq \chi_{1,{\rm max}}$ and $0\leq \chi_2 \leq \chi_{2,{\rm max}}$.
Volumetric merger rates evolve as a power law in $(1+z)$, such that
    \begin{equation}
    p(z|\Lambda_\mathrm{inj}) \propto \frac{1}{1+z} \frac{dV_c}{dz} (1+z)^\kappa \qquad (z\leq z_\mathrm{max}),
    \label{eq:p-z}
    \end{equation}
where $dV_c/dz$ is the differential comoving volume per unit redshift.

As described in the main text, we supplement the LIGO-Virgo-KAGRA pipeline injections with additional batches of ``hopeless'' events that are confidently undetectable, and ``certain'' events that are guaranteed to be detected if one or more LIGO instrument is in observing mode.
The distributions of these hopeless and certain injection sets closely follow the LIGO-Virgo-KAGRA pipeline injections, but are chosen to have a shallower primary mass distribution (and, in the BBH case, shallower growth of the merger rate with redshift) in order to yield training data that more uniformly covers the compact binary parameter space.
We additionally produced auxiliary hopeless injections spanning a broad range of masses and redshifts.

The specific hyperparameters characterizing each of these injection sets are detailed in Table~\ref{tab:inj-population}.
In addition to hyperparameter values, this table also indicates which convention, Eq.~\eqref{eq:p-m2-conditional} or Eq.~\eqref{eq:p-m2-not}, is followed when defining a secondary mass distribution.
The penultimate column indicates the reference frequencies at which component spins were defined; these differ slightly between injection sets.
The final column indicates the number of events drawn from each set for use as training data.

We note that the total number of available pipeline injections is larger than the numbers we used for training.
As discussed in Sec.~\ref{sec:loss}, we introduce additional terms in the loss function involving the integrated detection efficiencies of several reference populations (see Eq.~\eqref{eq:xi-loss} and Appendix~\ref{app:reference-populations} below).
These terms terms are slow to evaluate.
The chosen number of injections yielded a good compromise between network accuracy and overall training time.
Additionally, the specific ratios in Table~\ref{tab:inj-population} between pipeline, hopeless, and certain injections were found to yield better network performance than when simply training with additional available pipeline injections.

\subsection{Reference populations for augmented training loss}
\label{app:reference-populations}

\begin{table*}[]
    \setlength{\tabcolsep}{3pt}
    \renewcommand{\arraystretch}{1.2}
    \centering
    \footnotesize
    \caption{
    Description of the reference populations used in augmented training loss function, as defined in Eq.~\eqref{eq:xi-loss} and surrounding text.
    Included in the table are the hyperparameters defining each reference population, the convention used in defining a secondary mass distribution, the true detection efficiency $\xi$ associated with each, and the number $N$ of random draws from each population used during training.
    }
    \begin{tabular}{r | l l l l l l l l l l l | l l}
    \hline \hline
    Reference Distribution 
        & $m_{1,{\rm min}}$
        & $m_{1,{\rm max}}$ 
        & $\alpha$
        & $m_{2,{\rm min}}$
        & $m_{2,{\rm max}}$ 
        & $\beta_q$ 
        & Convention
        & $\chi_{1,{\rm max}}$ 
        & $\chi_{2,{\rm max}}$
        & $\kappa$ 
        & $z_{\rm max}$ 
        & $\xi$ 
        & $N$ \\
    \hline
    BBH (``Astrophysical'')
        & $5\,M_\odot$
        & $100\,M_\odot$ 
        & $-3$
        & $2\,M_\odot$
        & $100\,M_\odot$ 
        & $1$ 
        & Eq.~\eqref{eq:p-m2-conditional}
        & $0.998$ 
        & $0.998$ 
        & $4$ 
        & $1.9$
        & $3.8\times10^{-4} $
        & $2\times10^5$ \\
    BBH (``Injection-like'')
        & $2\,M_\odot$
        & $100\,M_\odot$ 
        & $-2.35$
        & $2\,M_\odot$
        & $100\,M_\odot$ 
        & $1$ 
        & Eq.~\eqref{eq:p-m2-conditional}
        & $0.998$ 
        & $0.998$ 
        & $1$ 
        & $1.9$
        & $1.1\times10^{-3} $
        & $2\times10^5$\\
    NSBH (``Injection-like'')
        & $2.5\,M_\odot$
        & $60\,M_\odot$ 
        & $-2.35$
        & $1\,M_\odot$
        & $2.5\,M_\odot$ 
        & $0$ 
        & Eq.~\eqref{eq:p-m2-not}
        & $0.998$ 
        & $0.4$ 
        & $0$ 
        & $0.25$
        & $1.1\times10^{-2}$
        & $10^4$ \\
    BNS (``Injection-like'')
        & $1\,M_\odot$
        & $2.5\,M_\odot$ 
        & $0$
        & $1\,M_\odot$
        & $2.5\,M_\odot$ 
        & $0$ 
        & Eq.~\eqref{eq:p-m2-not}
        & $0.4$ 
        & $0.4$ 
        & $0$ 
        & $0.15$        
        & $1.6\times10^{-2}$
        & $10^4$  \\
    \hline
    \hline
    \end{tabular}
    \label{tab:reference-population}
\end{table*}

When describing our training loss function in Sec.~\ref{sec:loss}, we introduced an additional regularization term [Eq.~\eqref{eq:xi-loss}] used to motivate the network to accurately recover integrated detection efficiencies.
When training our emulator, we sum Eq.~\eqref{eq:xi-loss} across four reference populations.
These reference populations followed the same functional forms used to define and generate training data; see Eqs.~\eqref{eq:p-m1} through \eqref{eq:p-z}.
The specific hyperparameters characterizing each are given in Table~\ref{tab:reference-population}.
Also listed are the target detection efficiencies $\xi_I$ as estimated using pipeline injections, the number $N_I$ of draws from each population used to estimate $\hat \xi_I$ at each training step.
The number of draws from each population were chosen to yield similar expected precisions $\sigma_{\xi}/\xi = 1/\sqrt{\xi_I N_I}$ for each population's detection efficiency [see Eq.~\eqref{eq:xi-likelihood}] while also managing local memory requirements.

\subsection{Network structure and ensemble training}
\label{app:more-training}

Input data are regularized via a linear transformation to the unit interval (\texttt{sklearn.StandardScaler}).
The input and hidden layers use a LeakyReLU activation function with a slope parameter of $10^{-3}$, while the final layer has a sigmoid activation function, rescaled to yield values on the interval $\{0,0.94\}$.
Initial neuron weights were randomly drawn from a zero-mean Gaussian distribution with standard deviation $0.01$ and biases  initially set to zero; the exception is the final output neuron, whose initial bias was set to $\ln(10^{-3})$.

These choices were made after experimenting with a large number of alternatives.
We found that the most impactful design choices are (\textit{i}) the explicit use of amplitude parameters [Eq.~\eqref{eq:amps}] as well as the polarization angle as input parameters, (\textit{ii}) the adoption of a relatively shallow but wide network, rather than a narrower network with more hidden layers, and (\textit{iii}) the precise number of additional ``certain'' and ``hopeless'' injections we use to augment the LIGO-Virgo-KAGRA injection sets.

\begin{figure}
    \centering
    \includegraphics[width=0.45\textwidth]{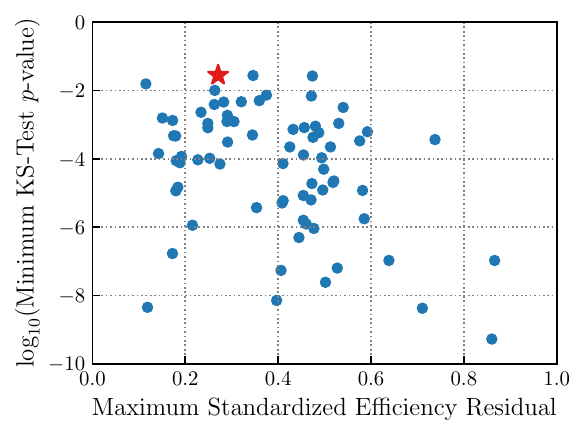}
    \caption{
    Summary statistics grading trained neural networks, as described in Appendix~\ref{app:more-training}.
    We trained an ensemble of networks, each with a different realization of training data and random initializations.
    The $y$-axis values indicate faithfulness in recovering correct parameter distributions of found compact binaries, while $x$-axis values indicate network accuracy in recovering integrated detection efficiencies.
    The fiducial network chosen in our study is indicated with a red star.
    }
    \label{fig:network-ensemble}
\end{figure}

After finalizing all details, we trained an ensemble of approximately 50 networks, each with random initialization conditions.
We graded the trained networks on two criteria.
First, for each network we computed predicted distributions of compact binary parameters and computed KS-test $p$-values between these predictions and the distributions actually recovered via pipeline injections (as in Figs.~\ref{fig:bbh}--\ref{fig:nsbh}).
For each network we record the minimum $p$-value, taken across all three source classes and all compact binary parameters.
Second, we compute predicted detection efficiencies for the reference populations listed in Table~\ref{tab:reference-population}, and, for each population, the standardized residual $(\hat \xi - \xi)/\sigma_\xi$ between the predicted and target values, where $\sigma_\xi = \sqrt{\xi/N}$ and $N$ is the total number of trials performed in the given computation [see Eq.~\eqref{eq:xi-likelihood}].
We record the maximum standardized residual for each network.
The results are shown in Fig.~\ref{fig:network-ensemble}, with each point representing a trained network from among the ensemble.
The fiducial network adopted for this paper is marked with a red star.

\section{Binary Black Hole Population Models}
\label{app:population-models}

\begin{table*}[]
    \setlength{\tabcolsep}{5pt}
    \renewcommand{\arraystretch}{1.2}
    \centering
    \caption{
    Hyperparameters specifying the binary black hole population models used throughout this work, as defined in Appendix~\ref{app:population-models}.
    The second column defines the distributions randomly sampled to obtain Fig.~\ref{fig:efficiency-check}.
    The third column gives the fixed values adopted when producing Fig.~\ref{fig:narrow-spins} ($\sigma_\chi$ is varied in this figure, and so is not given a value below).
    Finally, the fourth column gives the priors adopted when hierarchically inferring the binary black hole population in Sec.~\ref{sec:hierarchical-inference} and Fig.~\ref{fig:inference-comparison}.
    }
    \begin{tabular}{r | l l l}
    \hline \hline
    \multirow{2}*{Parameter}
        & Detection Efficiencies  
        & Dynamic injection regeneration
        & Hierarchical Inference \\[-3pt]
    {}
        & \textit{(Fig.~\ref{fig:efficiency-check})} 
        & \textit{(Fig.~\ref{fig:narrow-spins})}
        & \textit{(Fig.~\ref{fig:inference-comparison})} \\
    \hline
    $\mu_m$
        & $\mathrm{U}(20\,M_\odot,50\,M_\odot)$
        & $35\,M_\odot$
        & $\mathrm{U}(20\,M_\odot,50\,M_\odot)$ \\
    $\sigma_m$
        & $\mathrm{U}(2\,M_\odot,15\,M_\odot)$
        & $5\,M_\odot$
        & $\mathrm{U}(2\,M_\odot,15\,M_\odot)$ \\
    $f_p$
        & $\mathrm{LU}(10^{-6},1)$ 
        & $10^{-3}$
        & $\mathrm{LU}(10^{-6},1)$ \\
    $\alpha$
        & $\mathrm{N}(-2,3)$ 
        & $-3$
        & $\mathrm{N}(-2,3)$ \\
    $m_\mathrm{low}$
        & $\mathrm{U}(5\,M_\odot,15\,M_\odot)$ 
        & $10\,M_\odot$
        & $\mathrm{U}(5\,M_\odot,15\,M_\odot)$ \\
    $\delta m_\mathrm{low}$
        & $\mathrm{LU}(0.1\,M_\odot,10\,M_\odot)$ 
        & $1\,M_\odot$
        & $\mathrm{LU}(0.1\,M_\odot,10\,M_\odot)$ \\
    $m_\mathrm{high}$ 
        & $\mathrm{U}(50\,M_\odot,100\,M_\odot)$ 
        & $80\,M_\odot$
        & $\mathrm{U}(50\,M_\odot,100\,M_\odot)$ \\
    $\delta m_\mathrm{high}$
        & $\mathrm{LU}(10^{0.5}\,M_\odot,10^{1.5}\,M_\odot)$
        & $10\,M_\odot$
        & $\mathrm{LU}(10^{0.5}\,M_\odot,10^{1.5}\,M_\odot)$ \\
    $\beta_q$ 
        & $\mathrm{N}(0,3)$ 
        & $2$
        & $\mathrm{N}(0,3)$ \\
    $\mu_\chi$
        & $\mathrm{U}(0,1)$ 
        & $0$
        & $\mathrm{U}(0,1)$ \\
    $\sigma_\chi$ 
        & $\mathrm{LU}(0.1,1)$ 
        & --
        & $\mathrm{LU}(0.1,1)$ \\
    $ f_\mathrm{iso}$ 
        & $\mathrm{U}(0,1)$ 
        & 0.5
        & $\mathrm{U}(0,1)$ \\
    $\mu_u$ 
        & $\mathrm{U}(-1,1)$ 
        & $1$
        & $\mathrm{U}(-1,1)$ \\
    $\sigma_u$
        & $\mathrm{U}(0.15,2.5)$ 
        & $0.5$
        & $\mathrm{U}(0.15,2.5)$ \\
    $\kappa$ 
        & $\mathrm{N}(0,5)$ 
        & $3$
        & $\mathrm{N}(0,5)$ \\
    \hline
    \hline
    \end{tabular}
    \label{tab:inference-population}
\end{table*}

In a number of places in the main text, we invoke or infer realistic models for the astrophysical population of binary black holes.
This includes the discussion of integrated detection efficiencies surrounding Fig.~\ref{fig:efficiency-check} in Sec.~\ref{sec:performance}, and in the population inference performed in Sec.~\ref{sec:hierarchical-inference}.
In this appendix we describe the precise population models used for these results.

The primary mass distribution of binary black holes is assumed to follow a mixture between a power-law distribution and a Gaussian peak, 
    \begin{equation}
    \begin{aligned}
     p(m_1) &\propto \mathfrak{t}(m_1) \Bigg[
     f_p \frac{e^{-(m_1 - \mu_m)^2/2\sigma_m^2}}{\sqrt{2 \pi \sigma_m^2}} \\
     & \quad + (1-f_p)\frac{(1+\alpha)\,m_1^\alpha}{(100\,M_\odot)^{1+\alpha} - (2\,M_\odot)^{1+\alpha}}
     \Bigg]
    \end{aligned}
    \label{eq:p-m1}
    \end{equation}
where $\mathfrak{t}(m_1)$ is a tapering function that sends the mass distribution to zero at sufficiently high and low masses:
    \begin{equation}
    \mathfrak{t}(m_1) \propto \begin{cases}
        e^{-(m_1-m_\mathrm{low})^2/2\,\delta m_\mathrm{low}^2} & (m_1< m_\mathrm{low}) \\
        1 & (m_\mathrm{low}\leq m_1 \leq m_\mathrm{high}) \\
        e^{-(m_1-m_\mathrm{high})^2/2\,\delta m_\mathrm{high}^2} & (m_1 > m_\mathrm{high}). \\
    \end{cases}
    \end{equation}
Secondary masses are assumed to follow a power law, conditioned on primary masses:
    \begin{equation}
    p(m_2|m_1) = \frac{(1+\beta_q)m_2^{\beta_q}}{m_1^{1+\beta_q} - (2\,M_\odot)^{1+\beta_q}}.
    \label{eq:p-m2-appendix}
    \end{equation}
Component spins are assumed to be independently and identically distributed, with spin magnitudes following a truncated Gaussian distribution on the interval $0\leq\chi\leq 1$,
    \begin{equation}
    p(\chi) = \sqrt\frac{2}{\pi\sigma_\chi^2} \frac{e^{-(\chi - \mu_\chi)^2/2\sigma_\chi^2}}{\mathrm{Erf}\left(\frac{1-\mu_\chi}{\sqrt{2 \sigma_\chi^2}}\right) + \mathrm{Erf}\left(\frac{\mu_\chi}{\sqrt{2\sigma_\chi^2}}\right)},
    \end{equation}
while cosine spin-orbit tilt angles $\theta$ are assumed to follow a mixture between a truncated Gaussian and an isotropic component,
    \begin{equation}
    \begin{aligned}
    & p(\cos\theta) = \frac{f_\mathrm{iso}}{2} \\
    & + (1-f_\mathrm{iso}) \sqrt\frac{2}{\pi\sigma_u^2} \frac{e^{-(\cos\theta - \mu_u)^2/2\sigma_u^2}}{\mathrm{Erf}\left(\frac{1-\mu_u}{\sqrt{2 \sigma_u^2}}\right) + \mathrm{Erf}\left(\frac{1 + \mu_u}{\sqrt{2\sigma_u^2}}\right)},
    \end{aligned}
    \end{equation}
defined on the interval $-1 \leq \cos\theta \leq 1$.
The black hole merger rate per unit volume is assumed to evolve as a power law in $1+z$, such that the total number of mergers per unit source-frame time $dt_s$ per unit comoving volume $dV_c$ is
    \begin{equation}
    \frac{dN}{dt_s dV_c}(z) \propto (1+z)^\kappa.
    \label{eq:R-z-appendix}
    \end{equation}
The corresponding probability distribution of binary black hole redshifts is
    \begin{equation}
    p(z) \propto \frac{dV_c}{dz} (1+z)^{\kappa-1},
    \label{eq:p-z}
    \end{equation}
where $dV_c/dz$ is the differential comoving volume per unit redshift and the additional factor of $1+z$ converts from source-frame to detector-frame time.

Table~\ref{tab:inference-population} gives the priors and/or values adopted for the hyperparameters describing Eqs.~\eqref{eq:p-m1}--\eqref{eq:p-z} at different points throughout the paper.
The second column shows the hyperparameter distributions sampled to obtain the results in Fig.~\ref{fig:efficiency-check}.
The third column gives the fixed values chosen for Fig.~\ref{fig:narrow-spins} (the value of $\sigma_\chi$ is left blank, as this parameter is varied), when demonstrating the dynamic regeneration of injections using the \pdettheta{} emulator.
And the final column gives the hyperpriors adopted when performing full hierarchical inference in Sec.~\ref{sec:hierarchical-inference}, both when using traditional injection reweighting and when instead leveraging the neural network emulator.

\section{Hierarchical Inference Methods and Hybrid Injection Generation}
\label{app:hierarchical-inference}

We perform hierarchical inference of the binary black hole population following standard methods, as described in, e.g., Refs.~\cite{Mandel2019,Vitale2020}.
Consider a set of detected gravitational-wave events, with data $\{d_I\}$, where $I \in [1,N_\mathrm{events}]$ indexes each event.
We have posterior samples $\{\theta_{I,j}\}$ on the properties of each event, generated according to some generic prior \ppe{}.
The likelihood that this data arose from a given compact binary population $\Lambda$ is, then,
    \begin{equation}
    p(\{d\}|\Lambda) \propto e^{-N_\mathrm{exp}(\Lambda)} \prod_{I=1}^{N_{\rm events}}
        \left\langle \frac{dN/d\theta(\theta_{I,j};\Lambda)}{p(\theta_{I,j}|\Lambda_\mathrm{pe})} \right\rangle_j,
    \label{eq:full-likelihood}
    \end{equation}
where $\langle \cdot \rangle_j$ denotes an ensemble average over the posterior samples $j$.

The factor $dN/d\theta(\theta;\Lambda)$ in Eq.~\eqref{eq:full-likelihood} is the predicted number density of compact binary mergers (e.g. number of events per unit redshift, per unit mass, etc.).
This is related to the volumetric source-frame rate by
    \begin{equation}
    \frac{dN}{d\theta} = \frac{T_{\rm obs}}{1+z} \frac{dV_c}{dz} \frac{dN}{dt_s\,dV_c\,dm_1\,dq\,...}
    \label{eq:obs-rate}
    \end{equation}
where $T_\mathrm{obs}$ is our experiment's total duration.
We define the source-frame rate following the models described in Appendix~\ref{app:population-models}.
In particular, it is convenient to parametrize this function as
    \begin{equation}
    \begin{aligned}
    &\frac{dN}{dt_s\,dV_c\,dm_1\,dq\,...}
        = R_{\rm ref} \left(\frac{1+z}{1+z_{\rm ref}}\right)^\kappa
            \frac{p(m_1)}{p(m_1 = m_\mathrm{ref})} \\[5pt]
        &\qquad \times p(q) p(\chi_1) p(\chi_2) p(\cos\theta_1) p(\cos\theta_2),
    \end{aligned}
    \label{eq:source-frame-rate}
    \end{equation}
which avoids numerical computation of the normalization coefficient of the mass distribution.
This implies that $R_{\rm ref}$ is defined to be the volumetric merger rate per unit mass, defined at $m_1 = m_\mathrm{ref}$ and $z=z_{\rm ref}$; we take $m_{\rm ref} = 20\,M_\odot$ and $z=0.2$.

The term $N_{\rm exp}(\Lambda)$, meanwhile, is the expected number of detections if the compact binary population were indeed described by $\Lambda$.
This is directly related to the detection efficiency, $N_{\rm exp}(\Lambda) = N(\Lambda) \xi(\Lambda)$, where $N(\Lambda)$ is the total number of events occurring in the observation period, as can be obtained by integrating $dN/d\theta$.
When correcting for selection effects using a fixed suite of pipeline injections, $N_{\rm exp}(\Lambda)$ can be more directly estimated via
    \begin{equation}
    N_{\rm exp}(\Lambda) = \frac{1}{N_{\rm total}} \sum_i \frac{dN/d\theta(\theta_i;\Lambda)}{\pinji},
    \end{equation}
where the index $i$ ranges over found injections, $N_{\rm total}$ is the total number of injections performed, and \pinj{} is the distribution from which the injections were drawn; compare to Eq.~\eqref{eq:xi-reweighting}.
As described in the main text, we can alternatively use a trained \pdettheta{} emulator to evaluate $N_\mathrm{exp}$.
Following the algorithm in Sec.~\ref{sec:injection-regen-algorithm}, we directly generate a set of compact binary parameters $\{\theta_i\}\sim p(\theta|\Lambda)$ drawn from the proposed population $\Lambda$.
Then the expected number of detections is
    \begin{equation}
    \begin{aligned}
    N_{\rm exp}(\Lambda)
        &= \int \frac{dN}{d\theta}(\theta;\Lambda) \pdettheta{}\, d\theta \\
        &= N(\Lambda) \int p(\theta|\Lambda) \pdettheta{}\, d\theta \\
        &\approx N(\Lambda) \left\langle \hatpdetthetai \right\rangle_i,
    \end{aligned}
    \end{equation}
replacing the integral of \pdettheta{} over the compact binary probability distribution with the ensemble average of our emulator, \hatpdettheta{}, across the samples drawn from \ppop{}.
The remaining term $N(\Lambda)$ is the expected total number of events (detected or otherwise); this is found by integrating Eq.~\eqref{eq:obs-rate}.
Plugging in our model for the source-frame rate density [Eq.~\eqref{eq:source-frame-rate}], this yields
    \begin{equation}
    N(\Lambda) = R_{\rm ref}
            \frac{\int p(m_1) dm_1}{p(m_1=m_\mathrm{ref})}
            \frac{\int dV_c/dz\,(1+z)^{\kappa-1} dz}{(1+z_{\rm ref})^\kappa}
    \label{eq:Ntot}
    \end{equation}

In practice, we find that the above approach does not always yield stable results.
Specifically, while the large majority of samples drawn from realistic populations will lie at high redshifts, the detection efficiency primarily depends on the small fraction that happen to lie at low redshift, yielding a high-variance estimate of the integrated detection efficiency.
To circumvent this, we modify the algorithm in Sec.~\ref{sec:injection-regen-algorithm} by, in Step 1, drawing a single, \textit{fixed} set of redshifts ${\bm z}\sim p(z|\kappa=\kappa_{\rm ref})$ from a reference redshift distribution defined by some $\kappa_{\rm ref}$.
We choose $\kappa_{\rm ref} = -1.5$ in order to guarantee a large number of injections situated at small redshifts.
All other parameters are dynamically drawn as usual following the algorithm in Sec.~\ref{sec:injection-regen-algorithm}.
Evaluation of $N_{\rm exp}(\Lambda)$ then proceeds as above, but with an extra reweighting term accompanying the ensemble average over injections.
Define $\tilde \theta$ as the set of all binary parameters \textit{except} redshift, and similarly $\tilde \Lambda$ to be all population hyperparameters except $\kappa$.
Then
    \begin{equation}
    \begin{aligned}
    &N_{\rm exp}(\Lambda) \\
        &\quad= N(\Lambda) \int p(\tilde\theta|\tilde \Lambda)\, p(z|\kappa)
            P(\mathrm{det}|\tilde\theta,z)\, d\tilde\theta\,dz \\
        &\quad= N(\Lambda) \int p(\tilde\theta|\tilde\Lambda)\, p(z|\kappa_{\rm ref}) \frac{p(z|\kappa)}{p(z|\kappa_{\rm ref})}
            P(\mathrm{det}|\tilde\theta,z)\, d\tilde\theta\,dz \\
        &\quad\approx N(\Lambda) \left\langle
            \frac{p(z_i|\kappa)}{p(z_i|\kappa_{\rm ref})} \hat P(\mathrm{det}|\tilde\theta_i,z_i)
            \right\rangle_i,
    \end{aligned}
    \end{equation}
where in the final line we are now taking the average over our hybrid injections drawn from the fixed redshift distribution $p(z|\kappa_{\rm ref})$.
Note that the probability distributions appearing in the reweighting factors must be properly normalized [although the normalization coefficient of $p(z_i|\kappa)$ conveniently cancels with the redshift integral appearing in Eq.~\eqref{eq:Ntot} for $N(\Lambda)$].

\begin{figure*}
    \centering
    \includegraphics[width=0.99\textwidth]{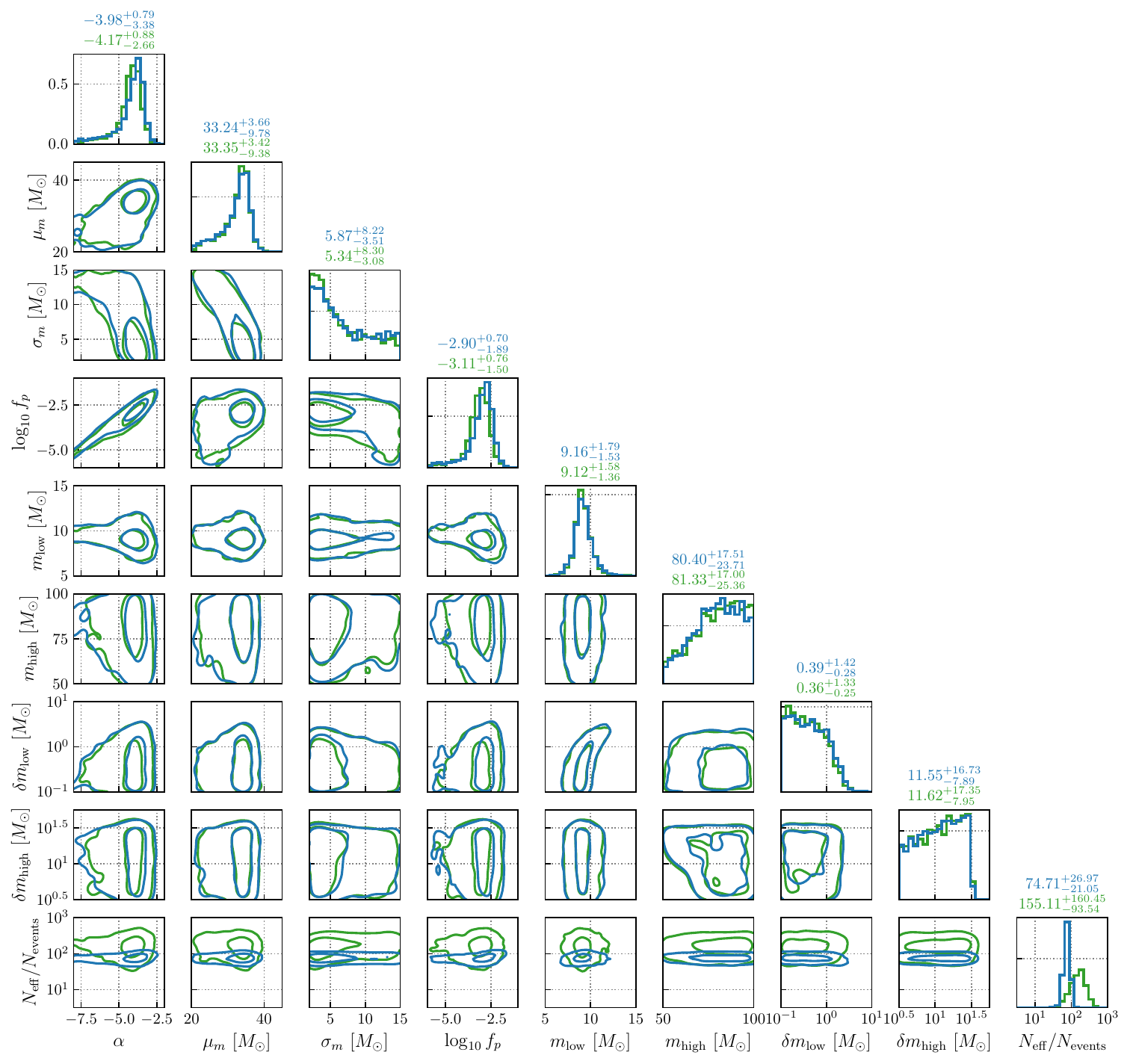}
    \caption{
    Inferred posteriors on the hyperparameters governing the binary black hole primary mass distribution, as discussed in Sec.~\ref{sec:hierarchical-inference}.
    Results using standard reweighting of pipeline injections are shown in green, while results obtained using our neural network $\pdettheta{}$ emulator are given in blue.
    The labels above each column give the median inferred hyperparameter values, with errors indicating central 95\% credible bounds.
    }
    \label{fig:corner-1}
\end{figure*}

\begin{figure*}
    \centering
    \includegraphics[width=0.75\textwidth]{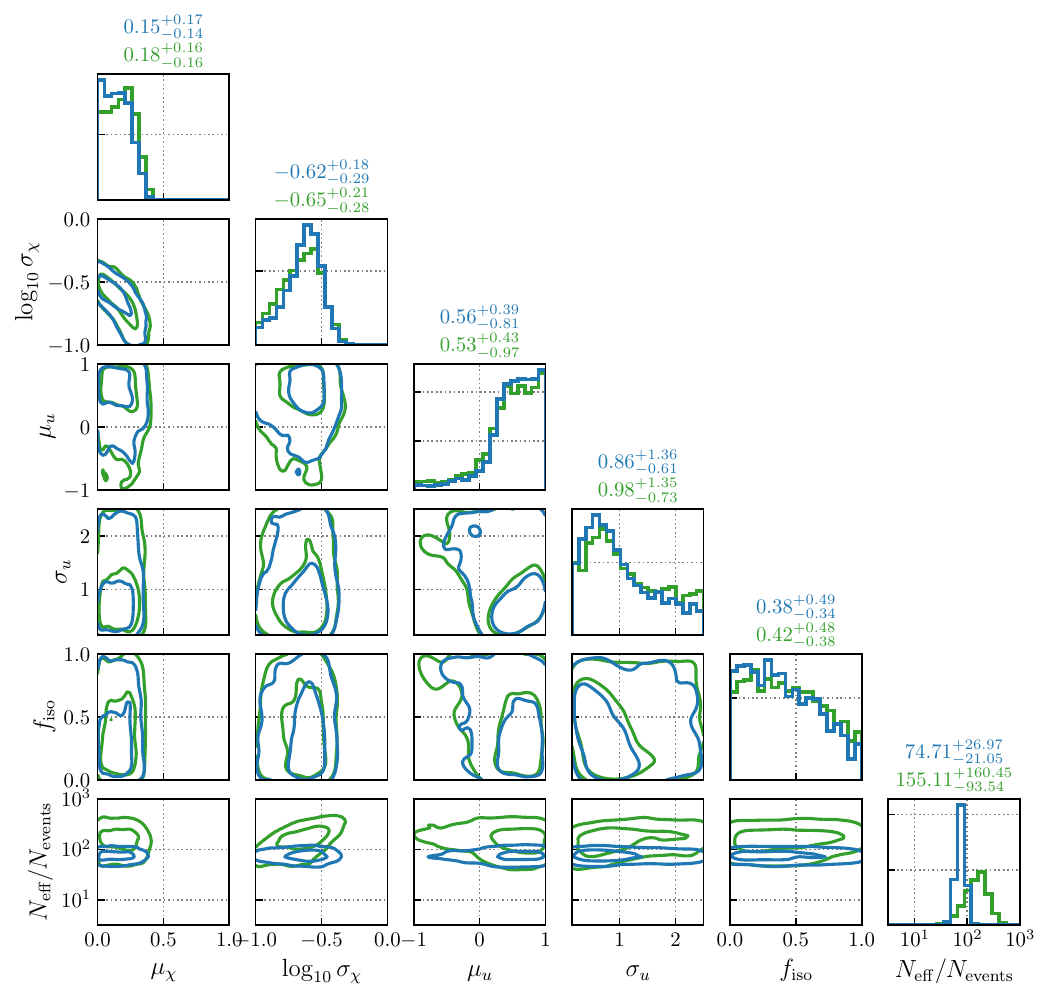}
    \caption{
    As in Fig.~\ref{fig:corner-1}, but showing hyperparameters associated with the binary black hole component spin distribution.
    }
    \label{fig:corner-2}
\end{figure*}

\begin{figure*}
    \centering
    \includegraphics[width=0.55\textwidth]{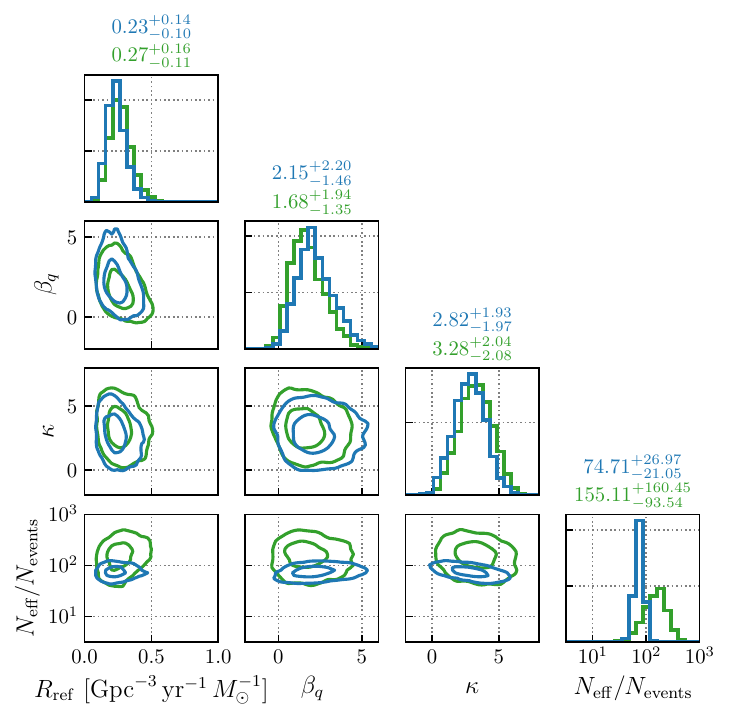}
    \caption{
    As in Fig.~\ref{fig:corner-1}, but showing hyperparameters governing the mass ratio distribution and redshift-dependent merger rate of binary black holes.
    Among all the hyperparameters across Figs.~\ref{fig:corner-1}--\ref{fig:corner-3}, $\beta_q$ and $\kappa$ [see Eqs.~\eqref{eq:p-m2-appendix} and \eqref{eq:R-z-appendix}] show the largest differences between both inference methods.
    When inferring the binary black hole population using our $\pdettheta{}$ emulator, we recover $\beta_q$ and $\kappa$ posteriors shifted towards larger and smaller values, respectively, relative to posteriors obtained through fixed injection reweighting.
    }
    \label{fig:corner-3}
\end{figure*}

We perform inference using the ``No U-Turn'' sampler~\cite{Hoffman2011} implemented in \texttt{numpyro}~\cite{numpyro1,numpyro2}, a probabilistic programming library built atop \texttt{jax}~\cite{jax}.
Our priors are as described in Appendix~\ref{app:population-models} and Table~\ref{tab:inference-population} therein.
We use posterior samples from the GWTC-2.1~\cite{GWTC2-1} and GWTC-3 catalogs~\cite{GWTC3}, made available via Zenodo~\cite{gwtc2-1-pe,gwtc3-pe}.
We specifically make use of the ``\texttt{C01:Mixed}'' samples, comprising results from a mixture of waveform models that all include the effects of spin precession and radiation from higher-order multipole moments.
For all inference runs, we monitor convergence by tracking the effective number of injections per catalog event [see Eq.~\eqref{eq:neff}], together with the effective number of posterior samples informing the likelihood assigned to each event.

\section{Additional results}
\label{app:more-results}

In the main text, we showed hierarchical inference results in the form of direct constraints on the distributions of black holes masses, spins, and redshifts.
In Figs.~\ref{fig:corner-1}--\ref{fig:corner-3} we alternatively show posteriors on the hyperparameters underlying these distributions.
As in the main text, green curves show posteriors obtained via traditional injection reweighting, while blue curves show results obtained using our neural network \pdettheta{} emulator.
Both sets of results are broadly consistent with on another, with the largest differences being slight shifts in the $\beta_q$ and $\kappa$ posteriors in Fig.~\ref{fig:corner-3}.
In each corner plot, we also show the effective number of injections (whether fixed injections used for reweighting or freshly-generated using the neural network) per event in the catalog sample, $N_{\rm eff}/N_{\rm events}$.
Hierarchical inference with injection reweighting yields a large range of effective injection counts, with values that are correlated with certain hyperparameters (e.g. $\alpha$, $\sigma_m$, and $\sigma_u$).
In contrast, using a \pdettheta{} emulator to draw from each target population of interest yields $N_{\rm eff}/N_{\rm events}$ values that are largely consistent across the full posterior and uncorrelated with specific hyperparameters.
The one exception is $\kappa$, which is slightly correlated with effective injection counts due to our use of the hybrid injection generation algorithm described above.

\bibliography{References}
\end{document}